\documentclass[twocolumn,showpacs,preprintnumbers,amsmath,amssymb]{revtex4}

\usepackage{graphicx}
\usepackage{dcolumn}
\usepackage{bm}

\newcommand{\Eq}[1]{Eq.~({\protect\ref{#1}})}

\newcommand{\Ref}[1]{Ref.\protect\cite{#1}}

\newcommand{\Sect}[1]{Sect.~\protect\ref{#1}}
\newcommand{\Appendix}[1]{Appendix.~\protect\ref{#1}}
\newcommand{\Fig}[1]{Fig.~\protect\ref{#1}}

\newcommand{\Table}[1]{Table~\protect\ref{#1}}

\newcommand{\Tate}{\rule{0cm}{1.1em}}
\newlength{\Tatescale}
\newcommand{\fslash}[1]{\mbox{$\!\not\!#1$}}

\newcommand{\Hs}{\hspace*{1em}}
\newcommand{\Bs}{\hspace*{-0.5em}}

\newcommand{\zr}[1]{\mbox{\hspace*{#1em}}}

\newcommand{\ZZ}{\mbox{\sf Z\zr{-0.45}Z}}

\newenvironment{@@@}{\mbox{ }\\{\bf @@@ begin @@@}\\}{\mbox{ }\\{\bf @@@ end @@@}\\}

\newlength{\figwidth}
\newcounter{subfigure}
\newcommand{\Cut}[1]{}
\begin{document}
\title{
Spin 3/2 Penta-quarks in anisotropic lattice QCD
}
\author{
  N.~Ishii$^1$ \footnote{E-mail : ishii@rarfaxp.riken.jp},
  T.~Doi$^2$ \footnote{E-mail: doi@quark.phy.bnl.gov},
  Y.~Nemoto$^3$ \footnote{E-mail: nemoto@hken.phys.nagoya-u.ac.jp},
  M.~Oka$^1$ \footnote{E-mail: oka@th.phys.titech.ac.jp},
  and H.~Suganuma$^4$ \footnote{E-mail: suganuma@th.phys.titech.ac.jp}
}
\affiliation{$^1$
  Department of Physics, H-27,
  Tokyo Institute of Technology,
  2-12-1 Oh-okayama,
  Meguro, Tokyo 152-8551, Japan
}
\affiliation{$^2$
  RIKEN BNL Research Center, Brookhaven National Laboratory,
  Upton, New York 11973, USA
}
\affiliation{$^3$
  Department of Physics, Nagoya University, Furo, Chikusa,
  Nagoya 464-8602, Japan
}
\affiliation{$^4$
  Department of Physics, Kyoto University, Kitashirakawaoiwake, 
  Kyoto 606-8502, Japan
}

\begin{abstract}
A high-precision  mass  measurement for the pentaquark (5Q) $\Theta^+$ in $J^{P}=3/2^{\pm}$ channel  
is performed in anisotropic quenched lattice QCD using a large number of gauge configurations as $N_{\rm conf}=1000$.   
We employ the standard Wilson gauge action  at $\beta=5.75$  and  the  $O(a)$ improved  Wilson (clover)  quark
action  with  $\kappa=0.1210(0.0010)0.1240$ on a  $12^3 \times 96$ lattice
with the  renormalized anisotropy  as $a_{\rm s}/a_{\rm t} =  4$.
The  Rarita-Schwinger  formalism  is  adopted  for  the  interpolating
fields.  Several types of  the interpolating fields with isospin $I=0$
are  examined  such as  (a)  the  NK$^*$-type,  (b) the  (color-)twisted
NK$^*$-type, (c) a diquark-type.
The chiral  extrapolation leads to only massive  states, i.e., $m_{\rm 5Q}
\simeq 2.1-2.2$ GeV in $J^P=3/2^-$ channel, and $m_{\rm 5Q} = 2.4-2.6$ GeV
in $J^P=3/2^+$ channel.
The analysis with the hybrid boundary condition(HBC) is performed
to investigate whether these states are compact 5Q resonances or not.
No low-lying compact 5Q resonance states are found below 2.1GeV.
\end{abstract}

\maketitle
\section{Introduction}
\label{introduction}
The recent discovery of  the manifestly exotic baryon $\Theta^+(1540)$
by the  LEPS group at SPring-8 has  made a great impact  on the exotic
hadron  physics  \cite{nakano}.  Apart  from  other pentaquark  baryon
candidates,   $\Xi^{--}(1862)$    \cite{NA49}   and   $\Theta_c(3099)$
\cite{H1}, several  other candidates of exotic hadrons  have also been
discovered, such as $X(3872)$, $D_s(2317)$, $S_0(3115)$, $X(3940)$ and
$Y(3840)$  \cite{exotics}.  They also  receive an  increasing interest
from theoretical side as well.
$\Theta^+(1540)$  is  supposed to  have  baryon  number $B=1$,  charge
$Q=+1$  and strangeness  $S=+1$. Since  $uudd\bar{s}$ is  the simplest
quark content to implement  this quantum number, $\Theta^+(1540)$ is a
manifestly exotic penta-quark(5Q) state.
The  penta-quark $\Theta^+$  had  been considered  several times  even
before the experimental discovery \cite{diakonov, jaffe-76, strottman,
weigel,praszalowicz}.
In particular,  \Ref{diakonov} provided  the direct motivation  of the
experimental search \cite{nakano}.
The  discovered peak  in  the  $nK^+$ invariant  mass  is centered  at
$1.54\pm 0.01$ GeV with a width smaller than 25 MeV.
At   the    present   stage,   some   groups    confirmed   the   LEPS
discovery\cite{diana,clas,saphir,experiments},    while   the   others
reported null results\cite{null}.
It will still take a while to establish the existence or non-existence
of $\Theta^+(1540)$ experimentally \cite{hicks}.
\Ref{saphir}  claims  that  $\Theta^+$  must be  isoscalar,  since  no
$\Theta^{++}$ is observed in the $pK^+$ invariant mass spectrum.

Enormous  theoretical   efforts  have  been  devoted   to  5Q  baryons
\cite{diakonov,jaffe-76,strottman,weigel,      praszalowicz,      oka,
zhu.review,    cohen,   itzhaki,    kim,   hosaka,    jaffe,   lipkin,
carlson-positive,  stancu,  jennings,  glozman,  enyo,  zhu,  matheus,
sugiyama,  carlson,  shinozaki,  huang, maezawa,  shlee,  narodetskii,
sugamoto,  suganuma,  okiharu,  bicudo, oset,  hosaka32,  nishikawa32,
takeuchi32, sugiyama32,  zhu32, inoue32, jaffe32,  huang32, capstic32,
dudek32,hyodo32,nam32}.
One of the most challenging problems in understanding its structure is
its extremely  narrow decay width  as $\Gamma \alt 1$  MeV \cite{pdg}.
Several ideas have been proposed:
(1) $I=2$  possibility \cite{capstic32},
(2) Jaffe-Wilczek's diquark picture \cite{jaffe},
(3) $\pi K N$ hepta-quark picture \cite{bicudo,kishimoto},
(4) string picture \cite{suganuma,sugamoto},
(5) $J^P=3/2^-$ possibility \cite{hosaka,enyo}.
Although each  gives a  mechanism to explain  the narrow  decay width,
none of them can satisfy  all the known properties of $\Theta^+(1540)$
simultaneously.
%

In  this  paper,  we   are  interested  in  the  $J=3/2$  possibility,
$J^P=3/2^-$ in particular.
Note that  the spin  of $\Theta^+(1540)$ has  not yet  been determined
experimentally.
%
In  the   constituent  quark  picture,  the  narrow   decay  width  of
$J^P=3/2^-$  penta-quarks  can  be  understood in  the  following  way
\cite{hosaka,enyo}.
We expect that  the special configuration $(0s)^5$ is  dominant in the
5Q ground-state in $J^P=3/2^-$ channel.
Although $J^P=3/2^-$ penta-quarks  can decay to KN in  the d-wave, the
spectroscopic factor to find d-wave KN states in the dominant $(0s)^5$
configuration vanishes.  Since the  decay is thus allowed only through
its sub-dominant d-wave configuration, the decay width is suppressed.
Note that it is further  suppressed by the d-wave centrifugal barrier,
leading  to  the  significantly  narrow  decay  width  of  $J^P=3/2^-$
penta-quarks.
%
A possible disadvantage of $J^P=3/2^-$ assignment is that such a state
tends  to be  massive due  to  the color-magnetic  interaction in  the
constituent quark models,
which seems to be one of the main reasons why there are only a limited
number  of   effective  model   studies  for  spin   3/2  penta-quarks
\cite{hosaka32,nishikawa32,takeuchi32,sugiyama32,zhu32,
inoue32,jaffe32,huang32,capstic32,hyodo32,nam32,enyo}.
However,  it is  not  clear whether  these  conventional framework  is
applicable  to a  new  exotic 5Q  system  as $\Theta^+(1540)$  without
involving any modifications.
%
Indeed,  a  model  was proposed  where  a  part  of  the role  of  the
color-magnetic   interaction  can   be  played   by   the  flavor-spin
interaction, which  makes the  mass-splitting between the  $1/2^-$ and
the $3/2^-$ states smaller \cite{takeuchi32}.

There  have been  several lattice  QCD  calculations of  5Q states  by
today\cite{scikor12,sasaki,chiu,kentacky,ishii12,rabbit,
lasscock12,alexandrou12,csikor122,holland,lasscock32}.  However, these
studies  are restricted to  $J^P=1/2^\pm$ channels  except for  a very
recent one \cite{lasscock32}.
Enormous efforts are being devoted to
more accurate  studies of $J^P=1/2^\pm$ states,  using the variational
technique to  extract multiple excited  states, among which  a compact
resonance state is sought for.
Indeed,  quite   large  scale  calculations  are   planned  and  being
performed\cite{fleming}  attempting to  elucidate  some of  the
mysterious natures  of $\Theta^+(1540)$ such as  its diquark structure
and/or non-localities desired in interpolating fields.
Here,   we  emphasize  again   that  these   studies  are   aiming  at
$J^P=1/2^\pm$ states, not at $3/2^\pm$ states.

In this paper, we present anisotropic lattice QCD results on 5Q states
in $J^P=3/2^\pm$ channels using a large number of gauge configurations
as  $N_{\rm conf}=1000$
as  an attempt to  search for  a low-lying  5Q state  in $J^P=3/2^\pm$
channel.
%
We  adopt the  standard Wilson  gauge  action at  $\beta=5.75$ on  the
$12^3\times 96$ lattice  with the renormalized anisotropy $a_{s}/a_{t}
= 4$.   The anisotropic lattice is  known to serve as  a powerful tool
for    high-precision    measurements    of    temporal    correlators
\cite{klassen,matsufuru,nemoto,ishii-gb}.
The large  number of gauge configurations $N_{\rm  conf}=1000$ plays a
key role in our calculation, because 5Q correlators in $J^P=3/2^{\pm}$
channels are found to be quite noisy.
For quark action, we adopt $O(a)$-improved Wilson (clover) action with
four  values  of  the  hopping parameter  as  $\kappa=0.1210  (0.0010)
0.1240$.
One of  the purpose of our  calculation is to examine  how the results
depend on the choice of interpolating field operators.
We  employ   several  types  of   interpolating  fields  as   (a)  the
NK$^*$-type,  (b) the (color-)twisted  NK$^*$-type, (c)  a diquark-type,
and adopt a smeared source to enhance the low-lying spectra.
%

In  $J^P=3/2^-$ channel,  we obtain  massive states  $m_{\rm 5Q}\simeq
2.1-2.2$ GeV  except for  the diquark-type interpolating  field, which
involves a considerable size  of the statistical error.
In  $J^P=3/2^+$  channel,  we   obtain  more  massive  states  $m_{\rm
5Q}\simeq 2.4-2.6$ GeV.
None of these  5Q states appear below the NK  threshold. Note that the
NK threshold  is raised up  by about $200-250$  MeV due to  the finite
extent of the spatial lattice as $L\simeq 2.15$ fm,
from which we expect the  penta-quark signal  to appear  below the
(raised)  NK  threshold  considering  the  empirical  mass  difference
between N+K(1440) and $\Theta^+(1540)$.
To  clarify whether our  5Q states  are compact  resonance or  not, we
perform an analysis with the hybrid boundary condition(HBC), which was
recently  proposed  in \Ref{ishii12}.
HBC analysis  indicates that no  compact 5Q resonance is  contained in
our 5Q states both in $J^P=3/2^\pm$ channels.

The paper is organized as follows.
In  \Sect{general.formalisms}, we discuss  the general  formalisms. We
begin   by  introducing   several  types   of   interpolating  fields,
determining their parity  transformation properties.  We next consider
the  temporal correlator  and its  spectral decomposition.  We finally
discuss the two-particle scattering states involved in 5Q spectra, and
introduce  the hybrid boundary  condition (HBC)  to examine  whether a
state of our concern is a compact resonance state or not.
\Sect{lattice}  is devoted to  the brief  descriptions of  our lattice
action and parameters.
In  \Sect{numerical.results},  we present  our  numerical results  for
$J^P=3/2^{\pm}$   channels   in   the   standard   periodic   boundary
condition(PBC).
We  show 5Q  correlators of  various interpolating  fields,  i.e., the
NK$^*$-type, the (color-)twisted NK$^*$-type, the diquark-type.
In  \Sect{HBC.analysis},  we attempt  to  determine  whether these  5Q
states  are compact  5Q  resonance states  or two-particle  scattering
states by using the HBC.
In \Sect{summary}, we summarize our results.

\section{General formalisms}
\label{general.formalisms}
\subsection{Interpolating fields}
We consider an iso-scalar  interpolating field of {\em NK$^*$-type} in
Rarita-Schwinger form \cite{ioffe,benmerrouche,hemmert} as
\begin{eqnarray}
  \psi_{\mu}
  &\equiv&
  \epsilon_{abc}
  \left( u^T_a C\gamma_5 d_b\right) u_c
  \cdot
  \left( \bar{s}_d \gamma_{\mu} d_d \right)
  \label{nkstar}
  \\\nonumber
  &&
  -
  \epsilon_{abc}
  \left( u^T_a C\gamma_5 d_b\right) d_c
  \cdot
  \left( \bar{s}_d \gamma_{\mu} u_d \right),
\end{eqnarray}
where  $\mu$ denotes  the  Lorentz  index, $a-d$  refer  to the  color
indices,  and  $C=\gamma_4\gamma_2$  denotes  the  charge  conjugation
matrix.
Unless otherwise indicated, the  gamma matrices are represented in the
Euclidean form given in \Ref{montvay}.
%

We  are  also  interested   in  the  {\em  (color-)twisted  NK$^*$-type}
interpolating field as
\begin{eqnarray}
  \psi_{\mu}
  &\equiv&
  \epsilon_{abc}
  \left( u^T_a C\gamma_5 d_b\right) u_d
  \cdot
  \left( \bar{s}_d \gamma_{\mu} d_c \right)
  \label{n8k8star}
  \\\nonumber
  &&
  -
  \epsilon_{abc}
  \left( u^T_a C\gamma_5 d_b\right) d_d
  \cdot
  \left( \bar{s}_d \gamma_{\mu} u_c \right),
\end{eqnarray}
which is an extension to the one originally proposed in \Ref{scikor12}
to study $J^P=1/2^P$ 5Q states.
It  has a  slightly more  elaborate color-structure  than \Eq{nkstar},
suggesting somewhat  stronger coupling  to a genuine  5Q state,  if it
exists, than simple NK$^*$ states.

Another   interpolating   fields  of   our   possible  interests   are
diquark-type interpolating fields such as
\begin{equation}
  \psi_{\mu}
  \equiv
  \epsilon_{abc}
  \epsilon_{def}
  \epsilon_{cfg}
  \left( u_a^T C\gamma_5 d_b \right)
  \left( u_d^T C\gamma_5 \gamma_{\mu} d_e \right)
  C\gamma_5 \bar s_g
  \label{diquark-type.SV}
\end{equation}
which    is     an    extension    to    the     one    proposed    in
Refs.~\cite{sasaki,sugiyama}.
The first  factor corresponds to  the scalar diquark (color  $\bar {\bf 3}$,
$I=0$, $J^P=0^+$), which is expected to play important roles in hadron
physics \cite{jaffe-exotica}.
The second factor  corresponds to the vector diquark  (color $\bar {\bf 3}$,
$I=0$,  $J^P=1^-$).
%
Note  that, although the  axial-vector diquark  (color $\bar {\bf 3}$, $I=1$,
$J^P=1^+$) is considered to play a more important role than the vector
diquark, it  cannot replace the  vector diquark due to  its iso-vector
nature.
%
Unless otherwise  indicated, we  refer to \Eq{diquark-type.SV}  as the
``{\em diquark-type}'' interpolating field.
We can also consider another interpolating field of diquark-type as
\begin{equation}
  \psi_{\mu}
  \equiv
  \epsilon_{abc}
  \epsilon_{def}
  \epsilon_{cfg}
  \left( u_a^T C d_b \right)
  \left( u_d^T C\gamma_5 \gamma_{\mu} d_e \right)
  C \bar s_g,
  \label{diquark-type.PV}
\end{equation}
which consists  of the pseudo-scalar  diquark (color $\bar  3$, $I=0$,
$J^P=0^-$)  and  the  vector  diquark.  However,  actual  lattice  QCD
calculation shows that  its correlator is afflicted with  quite a huge
statistical  error.
%
A possible reason  could be attributed to the fact  that both of these
diquark fields do not survive the non-relativistic limit.
Hence, we do not consider this interpolating field in this paper.

Under the spatial  reflection of the quark fields  as
\begin{equation}
  q(\tau,\vec x) \to \gamma_4 q(\tau,-\vec x),
\end{equation}
all of these interpolating fields transform as
\begin{equation}
  \psi_{i}(\tau,\vec x)
  \to
  -\gamma_4 \psi_{i}(\tau,-\vec x),
  \label{parity}
\end{equation}
for $i=1,2,3$.

\subsection{5Q correlators and parity projection}
We consider the Euclidean temporal correlator as
\begin{equation}
  G_{\mu\nu}(\tau)
  \equiv
  \sum_{\vec x}
  \left\langle
  \psi_{\mu}(\tau,\vec x) \bar\psi_{\nu}(0,\vec 0)
  \right\rangle,
  \label{correlator}
\end{equation}
where  $\sum_{\vec x}$  projects the  total 5Q  momentum to  zero.
Since  the  spin  3/2  contribution  from the  temporal  component  of
Rarita-Schwinger spinor  vanishes in the  rest frame, we  can restrict
ourselves   to  the   spatial  parts,   i.e.,   $\mu\nu=1,2,3$.   Now,
\Eq{correlator} is decomposed in the following way:
\begin{eqnarray}
  G_{ij}(\tau)
  &=&
  {\bf P}^{(3/2)}_{ij} G^{(3/2)}(\tau)
  +
  {\bf P}^{(1/2)}_{ij} G^{(1/2)}(\tau),
  \label{correlator.1}
\end{eqnarray}
where  $i,j=1,2,3$ denote  the spatial  part of  the  Lorentz indices,
${\bf P}^{(3/2)}$ and ${\bf P}^{(1/2)}$ denote the projection matrices
onto the spin 3/2 and 1/2 subspaces defined as
\begin{eqnarray}
  {\bf P}^{(3/2)}_{ij}
  &\equiv&
  \delta_{ij}   -  (1/3)\gamma_i\gamma_j,
  \\\nonumber
  {\bf P}^{(1/2)}_{ij}
  &\equiv&
  (1/3)\gamma_i\gamma_j.
\end{eqnarray}
They satisfy the following relations as
\begin{eqnarray}
  {\bf P}^{(3/2)}_{ij} {\bf P}^{(3/2)}_{jk} &=& {\bf P}^{(3/2)}_{ik}
  \\\nonumber
  {\bf P}^{(1/2)}_{ij} {\bf P}^{(1/2)}_{jk} &=& {\bf P}^{(1/2)}_{ik}
  \\\nonumber
  {\bf P}^{(1/2)}_{ij} + {\bf P}^{(3/2)}_{ij} &=& \delta_{ij}
  \\\nonumber
  {\bf P}^{(1/2)}_{ij} {\bf P}^{(3/2)}_{jk} &=& {\bf P}^{(3/2)}_{ij} {\bf P}^{(1/2)}_{jk} = 0.
\end{eqnarray}
Here, summations over repeated indices are understood.
$G^{(3/2)}(\tau)$ and $G^{(1/2)}(\tau)$  in \Eq{correlator} denote the
spin 3/2  and 1/2  contributions to $G(\tau)$,  respectively,
which  can  be  derived  by  operating  ${\bf  P}^{(3/2)}$  and  ${\bf
P}^{(1/2)}$ on $G(\tau)$, respectively.
(In  our   practical  lattice   QCD  calculation,  we   construct  the
Rarita-Schwinger   correlator   $G_{ij}(\tau)$   for   $i=1,2,3$   and
$j=3$(fixed), and  multiply ${\bf P}^{(3/2)}$ from the  left to obtain
$G^{(3/2)}(\tau)$.)

In the  asymptotic region  ($0 \ll \tau  \ll N_t$),  contaminations of
excited states are  suppressed.  Considering the parity transformation
property  \Eq{parity},  $G^{(3/2)}(\tau)$  and  $G^{(1/2)}(\tau)$  are
expressed in this region as\\
\begin{eqnarray}
  \lefteqn{G^{(3/2)}(\tau)}
  \label{correlator.2}
  \\\nonumber
  &=&
  P_+
  \left\{
  |\lambda_{3/2^-}|^2 e^{-\tau m_{3/2^-}}
  + |\lambda_{3/2^+}|^2 e^{-(N_t - \tau) m_{3/2^+}}
  \right\}
  \\\nonumber
  &-&
  P_-
  \left\{
  |\lambda_{3/2^+}|^2 e^{-\tau m_{3/2^+}}
  + |\lambda_{3/2^-}|^2 e^{-(N_t - \tau) m_{3/2^-}}
  \right\}
  \\\nonumber
  \lefteqn{G^{(1/2)}(\tau)}
  \\\nonumber
  &=&
  P_+
  \left\{
  |\lambda_{1/2^-}|^2 e^{-\tau m_{1/2^-}}
  + |\lambda_{1/2^+}|^2 e^{-(N_t - \tau) m_{1/2^+}}
  \right\}
  \\\nonumber
  &-&
  P_-
  \left\{
  |\lambda_{1/2^+}|^2 e^{-\tau m_{1/2^+}}
  + |\lambda_{1/2^-}|^2 e^{-(N_t - \tau) m_{1/2^-}}
  \right\},
\end{eqnarray}
where $P_{\pm}\equiv (1\pm\gamma_4)/2$  denote the projection matrices
onto the ``upper'' and ``lower'' Dirac subspaces, respectively.
$m_{3/2^\pm}$  and  $m_{1/2^\pm}$ denote  the  lowest-lying masses  in
$J^P=3/2^\pm$ and $1/2^\pm$ channels, respectively.
$\lambda_{3/2^\pm}$ and $\lambda_{1/2^\pm}$ represent the couplings to
the  interpolating field  \Eq{nkstar} with  $J^P=3/2^\pm$  and $1/2^\pm$
states, respectively.
In  \Eq{correlator.2}, we adopt  the anti-periodic  boundary condition
along the temporal direction.
A  brief  derivation  of  \Eq{correlator.1} and  \Eq{correlator.2}  is
presented in \Appendix{spectral.representation}.
The  forward propagation  is dominant  in the  region $0  <  \tau \alt
N_t/2$,  while the  backward  propagation is  dominant  in the  region
$N_t/2 \alt \tau < N_t$.
To separate  the negative (positive) parity  contribution, we restrict
ourselves  to  the region  $0  < \tau  \alt  N_t/2$,  and examine  the
``upper'' (``lower'') Dirac component.

\subsection{Scattering states involved in 5Q spectrum}
We  consider  the  (two-particle)  scattering states  involved  in  5Q
spectrum.   For $J^P=3/2^\pm$ iso-scalar  penta-quarks, NK  and NK$^*$
scattering states  play an important role. ($\Delta$K  does not couple
to the iso-scalar channel.)  These states are expressed as
\begin{equation}
  |N(\vec p,s)K(-\vec p)\rangle,
  \Hs
  |N(\vec p,s)K^*(-\vec p,i)\rangle,
\end{equation}
where $s$ and $i$ denote the  spin of the nucleon and K$^*$, and $\vec
p$ denotes the spatial momentum  allowed for a particular choice of the
spatial boundary condition adopted. For instance, if these hadrons are
subject to  the spatially  periodic boundary condition,  their momenta
are quantized as
\begin{equation}
  p_i = 2 n_i \pi /L, \Hs n_i \in \ZZ,
  \label{pbc}
\end{equation}
where $L$ denotes  the spatial extent of the  lattice. In contrast, if
they  are subject  to the  spatially anti-periodic  boundary condition,
their momenta are quantized as
\begin{equation}
  p_i = (2 n_i + 1)\pi /L, \Hs n_i \in \ZZ.
  \label{apbc}
\end{equation}
We first perform the parity projections. The positive and the negative
parity states are obtained in the following way:
\begin{eqnarray}
  \lefteqn{
    |NK(\pm)\rangle
  }
  \label{parity.projections1}
  \\\nonumber
  &=&
  |N(\vec p,s)K(-\vec p)\rangle
  \mp
  |N(-\vec p,s)K(\vec p)\rangle
  \\
  \lefteqn{
    |NK^*(\pm)\rangle
  }
  \label{parity.projections2}
  \\\nonumber
  &=&
  |N(\vec p,s)K^*(-\vec p,i)\rangle
  \mp
  |N(-\vec p,s)K^*(\vec p,i)\rangle.
\end{eqnarray}
Assuming that the interactions between N and K and between N and K$^*$
are weak, their energies are approximated as
\begin{eqnarray}
  E_{NK}
  &\simeq&
  \sqrt{m_N^2 + \vec p^2} + \sqrt{m_K^2 + \vec p^2}
  \\
  E_{N^*K}
  &\simeq&
  \sqrt{m_N^2 + \vec p^2} + \sqrt{m_{K^*}^2 + \vec p^2},
\end{eqnarray}
respectively.
The scattering  states which couple to  $J^P=3/2^\pm$ penta-quarks are
obtained  as  spin-3/2  projections  of  \Eq{parity.projections1}  and
\Eq{parity.projections2}.  The d-wave NK  states and the s-wave NK$^*$
states  can couple  to the  $J^P=3/2^-$ channel,  while the  p-wave NK
states and NK$^*$ states can couple to the $J^P=3/2^+$ channel.

The scattering states  with vanishing spatial momentum $\vec  p = \vec
0$ are exceptional in the following sense.
On the one hand, the  positive parity states vanish, because the first
terms coincides with the  second terms in \Eq{parity.projections1} and
\Eq{parity.projections2}  in the  right hand side.
On the  other hand,  the negative parity  states are  constructed only
from the spin degrees of freedom, i.e., the spin degrees of freedom of
the  nucleon  in \Eq{parity.projections1},  and  the  spin degrees  of
freedoms of the nucleon and K$^*$ in \Eq{parity.projections2}.
By   counting  the  degeneracy   of  the   resulting  states,   it  is
straightforward  to see  that no  d-wave states  are  contained, i.e.,
\Eq{parity.projections1}  gives only s-wave  NK states  in $J^P=1/2^-$
channel,  and that \Eq{parity.projections2}  gives only  s-wave NK$^*$
states in $J^P=1/2^-$ and $3/2^-$ channels.
%

\subsection{Hybrid boundary condition(HBC)}
\begin{table}
\begin{ruledtabular}
\begin{tabular}{lcccccc}
$\kappa$ & & 0.1210 & 0.1220 & 0.1230 & 0.1240 & emp. \\
\hline
NK$^*$(s-wave) &	 PBC  & 2.996  & 2.815  & 2.633  & 2.445  & 1.830  \\
NK$^*$(p-wave) &	 PBC  & 3.222  & 3.052  & 2.883  & 2.710  & 2.163  \\
NK(p/d-wave) &  	 PBC  & 2.987  & 2.806  & 2.624  & 2.438  & 1.865  \\
NK$^*$(s/p-wave) &	 HBC  & 3.167  & 2.995  & 2.823  & 2.647  & 2.084  \\
NK(p/d-wave) &  	 HBC  & 2.924  & 2.739  & 2.553  & 2.363  & 1.770  \\
\end{tabular}
\end{ruledtabular}
\caption{  Numerical  values of  NK  and  NK$^*$  thresholds for  each
hopping parameter  $\kappa$ in  the physical unit  GeV in  the spatial
lattice of the size $L \simeq 2.15$ fm in PBC and HBC.
The rightmost column labeled as ``emp.''  shows the thresholds for the
physical  values  of  N,  K,  and K$^*$  as  $m_{N}\simeq  0.94$  GeV,
$m_{K}\simeq 0.5$ GeV, $m_{K^*}\simeq 0.89$ GeV.
}
\label{thresholds}
\end{table}
In order to  determine whether a state of our concern  is a compact 5Q
resonance state  or a  scattering state of  two particles, we  use two
distinct  spatial   boundary  conditions(BC),  i.e.,   the  (standard)
periodic BC(PBC) and the hybrid BC(HBC), which is recently proposed in
\Ref{ishii12}.
In PBC, one imposes the spatially periodic BC on u,d and s-quarks.  As
a result, all the hadrons are subject to the periodic BC.
In  this case, due  to \Eq{pbc},  
all hadrons can take zero-momentum, 
and the smallest  non-vanishing momentum
$\vec p_{\rm min}$ is of the form as
\begin{equation}
  (\pm 2\pi/L,0,0),\Hs
  (0,\pm 2\pi/L,0),\Hs
  (0,0,\pm 2\pi/L),
\end{equation}
which gives 
\begin{equation}
  |\vec p_{\rm min}^{\rm \,PBC}| = 2\pi/L.
  \label{pmin.pbc}
\end{equation}
On the other hand, in HBC, we impose the spatially anti-periodic BC on
u  and d-quarks,  whereas  the  spatially periodic  BC  is imposed  on
s-quark.      Since    N($uud,udd$),     K($u\bar{s},d\bar{s}$)    and
K$^*$($u\bar{s},d\bar{s}$) contain odd numbers of u and d quarks, they
are subject to the anti-periodic BC.
Therefore, due  to \Eq{apbc}, N, K  and K$^*$ cannot  have a vanishing
momentum in HBC. The smallest  possible momentum $\vec p_{\rm min}$ is
of the form as
\begin{equation}
  (\pm \pi/L,\pm \pi/L, \pm \pi/L).
\end{equation}
Hence, its norm $|\vec p_{\rm min}|$ is expressed as
\begin{equation}
  |\vec p_{\rm min}^{\rm \,HBC}| = \sqrt{3}\pi/L.
  \label{pmin.hbc}
\end{equation}
In  contrast,  $\Theta^+$($uudd\bar{s}$) is  subject  to the  spatially
periodic BC,  since it contains even  number of u and  d quarks.
Therefore, $\Theta^+$ can have the vanishing momentum.

Switching from PBC, HBC  affects the low-lying two-particle scattering
spectrum.  A drastic change is expected in the s-wave NK$^*$ channel.
In PBC, the energy of the lowest NK$^*$ state is given as
\begin{equation}
  E_{\rm min}^{\rm PBC}(NK^*(\mbox{s-wave}))
  \simeq
  m_N + m_{K^*}.
\end{equation}
In  contrast, in  HBC, since  both N  and K$^*$  are required  to have
non-vanishing momenta $|\vec p_{\rm min}| = \sqrt{3}\pi/L$, the energy of
the lowest NK$^*$ state is raised up as
\begin{eqnarray}
  \lefteqn{
    E_{\rm min}^{\rm HBC}(NK^*(\mbox{s-wave}))
  }
  \\\nonumber
  &\simeq&
  \sqrt{m_N^2 + 3\pi^2/L^2} + \sqrt{m_{K^*}^2 + 3\pi^2/L^2}.
\end{eqnarray}
Note that the shift amounts typically  to a few hundred MeV for $L\sim
2$ fm.

HBC affects  NK(d-wave), NK(p-wave), NK$^*$(p-wave)  as well. However,
these changes  are not as  drastic as that in  NK$^*$(s-wave), because
they  are induced by  the minor  change in  the minimum  momentum from
$|\vec p_{\rm min}|=2\pi/L$ to $\sqrt{3}\pi/L$.
In PBC, the energies of  the lowest two-particles states are expressed
as
\begin{eqnarray}
  \lefteqn{
    E_{\rm min}^{\rm PBC}(NK(\mbox{p/d-wave}))
  }
  \\\nonumber
  &\simeq&
  \sqrt{m_N^2 + 4\pi^2/L^2} + \sqrt{m_{K}^2 + 4\pi^2/L^2}
  \\\nonumber
  \lefteqn{
    E_{\rm min}^{\rm PBC}(NK^*(\mbox{p-wave}))
  }
  \\\nonumber
  &\simeq&
  \sqrt{m_N^2 + 4\pi^2/L^2} + \sqrt{m_{K^*}^2 + 4\pi^2/L^2}.
\end{eqnarray}
In HBC, they are shifted down as
\begin{eqnarray}
  \lefteqn{
    E_{\rm min}^{\rm HBC}(NK(\mbox{p/d-wave}))
  }
  \\\nonumber
  &\simeq&
  \sqrt{m_N^2 + 3\pi^2/L^2} + \sqrt{m_{K}^2 + 3\pi^2/L^2}
  \\\nonumber
  \lefteqn{
    E_{\rm min}^{\rm HBC}(NK^*(\mbox{p-wave}))
  }
  \\\nonumber
  &\simeq&
  \sqrt{m_N^2 + 3\pi^2/L^2} + \sqrt{m_{K^*}^2 + 3\pi^2/L^2}.
\end{eqnarray}
Numerical  values  of  NK  and  NK$^*$  thresholds  for  each  hopping
parameter in  spatial lattice of the  size $L\simeq 2.15$  fm for both
PBC and HBC are summarized in \Table{thresholds}.

In contrast to the scattering states,  HBC is not expected to affect a
compact 5Q resonance $\Theta^+$ so much.
Since $\Theta^+(uudd\bar{s})$ can have vanishing momentum also in HBC,
the shift  of the penta-quark  mass $m_{\rm 5Q}$ originates  only from
the change  in its  intrinsic structure.  In  this case, the  shift is
expected  to  be  less  significant  than the  shift  induced  by  the
kinematic reason as is the case in N, K, and K$^*$.
Now our way to find a compact 5Q resonance state is to seek for such a
state which is not affected by HBC.

\section{Lattice actions and parameters}
\label{lattice}
\begin{table*}
\begin{ruledtabular}
\begin{tabular}{ccccccccccl}
$\beta$ & $\gamma_{\rm  G}$ & $a_{\rm s}/a_{\rm t}$ & $a_{s}^{-1}$  [GeV] & Size &
$N_{\rm conf}$ & $u_{s}$ & $u_{t}$ & $\gamma_F$ & $\kappa_c$ &
Values of $\kappa$ \\
\hline
5.75 &  3.2552 & 4 &  1.100(6) & $12^3\times  96$ & 1000 &  0.7620(2) &
0.9871(0) & 3.909 & 0.12640(5) & 0.1240, 0.1230, 0.1220, 0.1210
\end{tabular}
\end{ruledtabular}
\caption{Parameters  of the lattice  simulation.  The  spatial lattice
spacing $a_{\sigma}$ is  determined with $r_0^{-1} = 385$  MeV for the
Sommer parameter.  The mean-field values of link variables ($u_\sigma$
and $u_\tau$) are defined in the Landau gauge.  $\kappa_c$ denotes the
critical  value of  $\kappa$. }
\label{table.lattice.parameters}
\end{table*}
To generate gauge field  configurations, we use the standard plaquette
action on the anisotropic lattice of the size $12^3\times 96$ as
\begin{eqnarray}
  S_{\rm G}
  &=&
  \frac{\beta}{N_c}
  \frac1{\gamma_{\rm G}}
  \sum_{x,i<j\le3}
  \mbox{Re} \mbox{Tr}
  \left\{ 1 - P_{ij}(x)\right\}
  \\\nonumber
  &+&
  \frac{\beta}{N_c}
  \gamma_{\rm G}
  \sum_{x,i\le 3}
  \mbox{Re} \mbox{Tr}
  \left\{ 1 - P_{i4}(x)\right\},
\end{eqnarray}
where $P_{\mu\nu}(x) \in  \mbox{SU(3)}$ denotes the plaquette operator
in  the  $\mu$-$\nu$-plane.   The   lattice  parameter  and  the  bare
anisotropy parameter are  fixed as $\beta \equiv 2N_c/g^2  = 5.75$ and
$\gamma_{\rm G}=3.2552$, respectively.  These values are determined to
reproduce the  renormalized anisotropy as $\xi\equiv  a_{s}/a_{t} = 4$
\cite{klassen}.
Adopting  the  pseudo-heat-bath  algorithm,  we pick  up  gauge  field
configurations every  500 sweeps after skipping 10,000  sweeps for the
thermalization.   We use  totally 1000  gauge field  configurations to
construct the temporal correlators.
Note  that  the  high  statistics  of  $N_{\rm  conf}=1000$  is  quite
essential  for our  study, because  the  5Q correlators  for spin  3/2
states are found to be rather noisy.
In  fact,   a  preliminary  analysis  with   less  statistics  $N_{\rm
conf}\simeq   500$   leads   to   a  spurious   resonance-like   state
\cite{ishii320}.
%
The  lattice spacing  is determined  from the  static  quark potential
adopting the Sommer parameter $r_0^{-1} = 395$ MeV ($r_0 \sim 0.5$ fm)
as $a_{s}^{-1} = 1.100(6)$ GeV ($a_{\rm s} \simeq 0.18$ fm). Note that
the lattice size $12^3\times  96$ amounts to $(2.15\mbox{fm})^3 \times
(4.30\mbox{fm})$ in the physical unit.

We adopt the $O(a)$-improved Wilson (clover) action on the anisotropic
lattice for quark fields $\psi$ and $\bar\psi$ as \cite{matsufuru}
\begin{eqnarray}
  S_{\rm F}
  &\equiv&
  \sum_{x,y} \bar\psi(x) K(x,y) \psi(y),
  \\\nonumber
  K(x,y)
  &\equiv&
  \renewcommand{\arraystretch}{1.8}
  \delta_{x,y}
  -
  \kappa_{\rm t}\left\{\Tate\right.
  \Bs
  \begin{array}[t]{l} \displaystyle
    (1 - \gamma_4)\; U_4(x)\; \delta_{x+\hat 4,y}
    \\\displaystyle
    +
    (1 + \gamma_4)\; U_4^\dagger(x - \hat 4)\; \delta_{x-\hat 4,y}
    \left.\Tate\right\}
  \end{array}
  \\\nonumber
  &-&
  \renewcommand{\arraystretch}{1.8}
  \kappa_{\rm s}
  \sum_i
  \left\{\Tate\right.
  \Bs
  \begin{array}[t]{l} \displaystyle
    (r - \gamma_i)\; U_i(x)\; \delta_{x+\hat i,y}
    \\\displaystyle
    +
    (r + \gamma_i)\; U_i^\dagger(x - \hat i)\; \delta_{x-\hat i,y}
    \left.\Tate \right\}
  \end{array}
  \\\nonumber
  &-&
  \kappa_{\rm s}\; c_E \sum_i \sigma_{i4} F_{i4} \delta_{x,y}
  -
  r\; \kappa_{s}\; c_B \sum_{i<j} \sigma_{ij} F_{ij} \delta_{x,y},
\end{eqnarray}
where  $\kappa_{\rm  s}$  and  $\kappa_{\rm  t}$  denote  the  hopping
parameters for the spatial  and the temporal directions, respectively.
The  field  strength  $F_{\mu\nu}$  is defined  through  the  standard
clover-leaf-type construction.
$r$  denotes the  Wilson  parameter. $c_{E}$  and  $c_{B}$ denote  the
clover coefficients.  
To achieve the tadpole improvement, the link variables are rescaled as
$U_i(x) \to U_i(x)/u_{\rm s}$ and $U_4(x) \to U_4(x)/u_{\rm t}$, where
$u_{\rm  s}$ and  $u_{\rm  t}$  denote the  mean-field  values of  the
spatial     and      temporal     link     variables,     respectively
\cite{matsufuru,nemoto}.
This is  equivalent to the  redefinition of the hopping  parameters as
the  tadpole-improved  ones  (with  tilde), i.e.,  $\kappa_{\rm  s}  =
\tilde\kappa_{\rm   s}   /   u_{\rm   s}$  and   $\kappa_{\rm   t}   =
\tilde\kappa_{\rm t} / u_{\rm t}$.
The   anisotropy   parameter    is   defined   as   $\gamma_F   \equiv
\tilde\kappa_{\rm t} / \tilde\kappa_{\rm s}$, which coincides with the
renormalized anisotropy $\xi = a_{\rm s} / a_{\rm t}$ for sufficiently
small quark mass at the tadpole-improved level \cite{matsufuru}.
For given $\kappa_{\rm s}$, the  four parameters $r$, $c_E$, $c_B$ and
$\kappa_{\rm s}/\kappa_{\rm t}$ should be, in principle, tuned so that
``Lorentz symmetry'' holds up to discretization errors of $O(a^2)$.
Here, $r$, $c_E$ and $c_B$  are fixed by adopting the tadpole improved
tree-level values as
\begin{equation}
  r   = \frac1{\xi},
  \Hs
  c_E =  \frac1{u_{\rm s} u_{\rm t}^2},
  \Hs
  c_B =  \frac1{u_{\rm s}^3}.
\end{equation}
Only the value  of $\kappa_{\rm s} / \kappa_{\rm  t} \left( = \gamma_F
\cdot (u_{\rm s}  / u_{\rm t}) \right)$ is  tuned nonperturbatively by
using the meson dispersion relation \cite{matsufuru}.
It is convenient to define $\kappa$ as
\begin{equation}
  \frac1{\kappa}
  \equiv
  \frac1{\tilde\kappa_{\rm s}}
  -
  2\left( \gamma_F - 3r - 4 \right).
\end{equation}
Then the bare  quark mass is expressed as  $m_0 = \frac1{2}(1/\kappa -
8)$ in the spatial lattice unit in the continuum limit.  This $\kappa$
plays the role of the  hopping parameter ``$\kappa$'' in the isotropic
formulation.
For  detail,  see  Refs.~\cite{nemoto,matsufuru},  where we  take  the
lattice  parameters.   The  values   of  the  lattice  parameters  are
summarized in \Table{table.lattice.parameters}.

\begin{table}
\begin{ruledtabular}
\begin{tabular}{llllll}
$\kappa$   & 0.1210 & 0.1220 & 0.1230 & 0.1240 & $\kappa_{\rm phys.}$\\
\hline
$m_{\pi}$  & 1.007(2) & 0.897(1) & 0.785(2) & 0.658(2)  & 0.140      \\
$m_{\rho}$ & 1.240(4) & 1.157(5) & 1.074(7) & 0.991(11) & 0.823(13)  \\
$m_{K}$    & 0.846(2) & 0.785(1) & 0.722(2) & 0.658(2)  & 0.476(2)   \\
$m_{K^*}$  & 1.119(6) & 1.076(7) & 1.033(9) & 0.991(11) & 0.902(15)  \\
$m_{N}$    & 1.877(4) & 1.739(3) & 1.600(4) & 1.454(5)  & 1.164(8)   \\
$m_{N^*}$  & 2.325(17) & 2.194(21)& 2.059(28) & 1.918(42) & 1.648(53)
\end{tabular}
\end{ruledtabular}
\caption{Masses  of $\pi$,  $\rho$, K,  K$^*$, N,  and N$^*$  for each
hopping  parameter $\kappa$  in the  physical unit  GeV.  $\kappa_{\rm
phys.}\simeq  0.1261$ denotes  the  value of  $\kappa$ which  achieves
$m_{\pi} \simeq 0.14$ GeV.}
\label{table.mass}
\end{table}
We  adopt four  values  of the  hopping  parameter as  $\kappa=0.1210,
0.1220, 0.1230$ and $0.1240$,  which correspond to $m_{\pi}/m_{\rho} =
0.81, 0.78, 0.73$ and $0.66$, respectively. These values roughly cover
the region $m_s \alt m \alt 2 m_s$.
For temporal direction, we  impose anti-periodic boundary condition on
all  the quark  fields.  For  spatial directions,  we  impose periodic
boundary condition on all  the quarks, unless otherwise indicated.  We
refer to this boundary condition as ``{periodic BC (PBC)}''.

By  keeping $\kappa_s=0.1240$  fixed for  $s$ quark,  and  by changing
$\kappa=0.1210-0.1240$ for  $u$ and $d$ quarks, we  perform the chiral
extrapolation to the physical quark mass region.
In the following part of the paper, we will use
\begin{equation}
  (\kappa_s,\kappa)=(0.1240,0.1220),
\label{typical.set}
\end{equation}
as a typical  set of hopping parameters in  presenting correlators and
effective mass plots.
For convenience, we summarize masses of $\pi$, $\rho$, K, K$^*$, N and
N$^*$($J^P=1/2^-$ baryon) for each hopping parameter $\kappa$ together
with their  values at the  physical quark mass  in \Table{table.mass}.
Here, the chiral extrapolations of  these particles are performed by a
linear function in $m_{\pi}^2$.
%
Unless  otherwise indicated,  we adopt  the jackknife  prescription to
estimate the statistical errors.

We  use a  smeared  source  to enhance  the  low-lying spectra.   More
precisely, we  employ spatially  extended interpolating fields  of the
gaussian size $\rho\simeq 0.4$ fm,  which is obtained by replacing the
quark fields  $q(x)$ in 5Q  interpolating fields by the  smeared quark
fields $q_{\rm smear}(x)$ in the Coulomb gauge as
\begin{equation}
  q_{\rm smear}(\tau, \vec x)
  \equiv
  {\cal N}
  \sum_{\vec y}
  \exp\left\{ - \frac{|\vec x - \vec y|^2}{2\rho^2} \right\}
  q(\tau, \vec y),
\label{eq.gaussian}
\end{equation}
where ${\cal N}$ is  an appropriate normalization factor.
%
%
For a practical use, we extend \Eq{eq.gaussian} appropriately so as to
fit a particular choice of the spatial boundary condition.
In  this paper, we  present correlators  with a  smeared source  and a
point sink.

\section{Numerical results on 5Q spectrum}
\label{numerical.results}
We  present our lattice  QCD results  on 5Q  spectrum in  the standard
periodic boundary condition(PBC) in this section.
\subsection{$J^P=3/2^-$ 5Q spectrum in PBC}
\begin{figure}
\begin{center}
\includegraphics[height=0.48\textwidth,angle=-90]{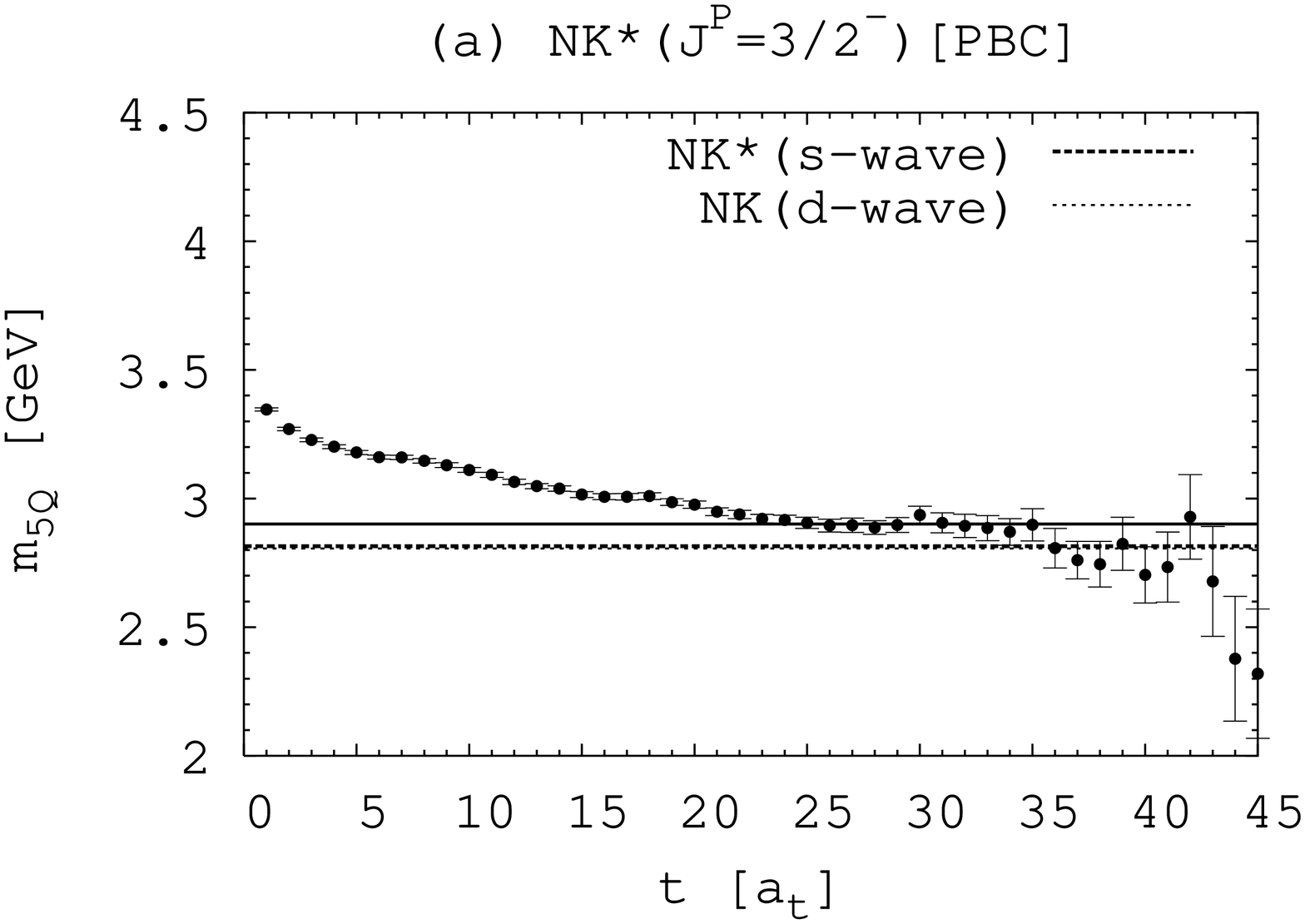}  
\\
\includegraphics[height=0.48\textwidth,angle=-90]{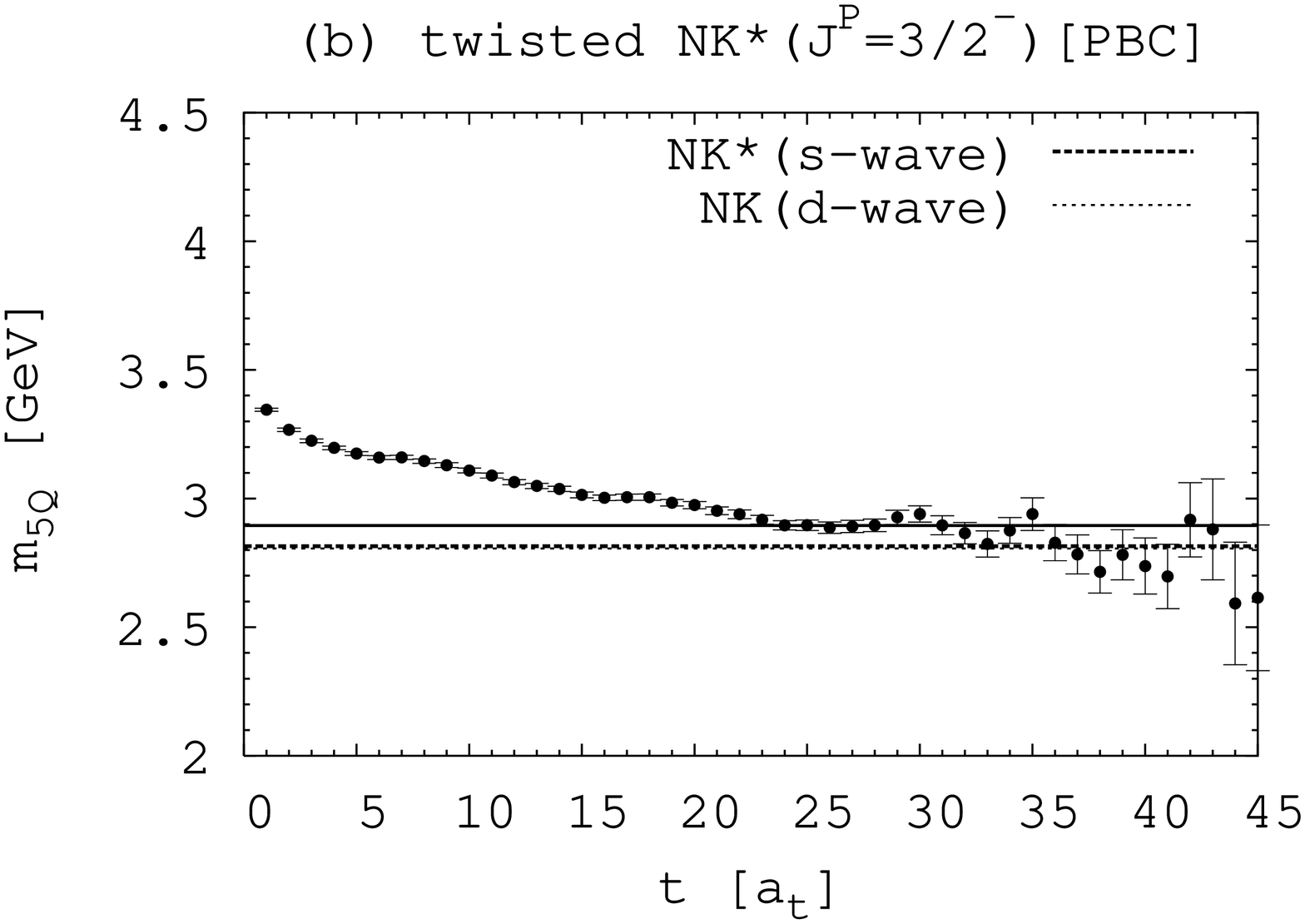}  
\\
\includegraphics[height=0.48\textwidth,angle=-90]{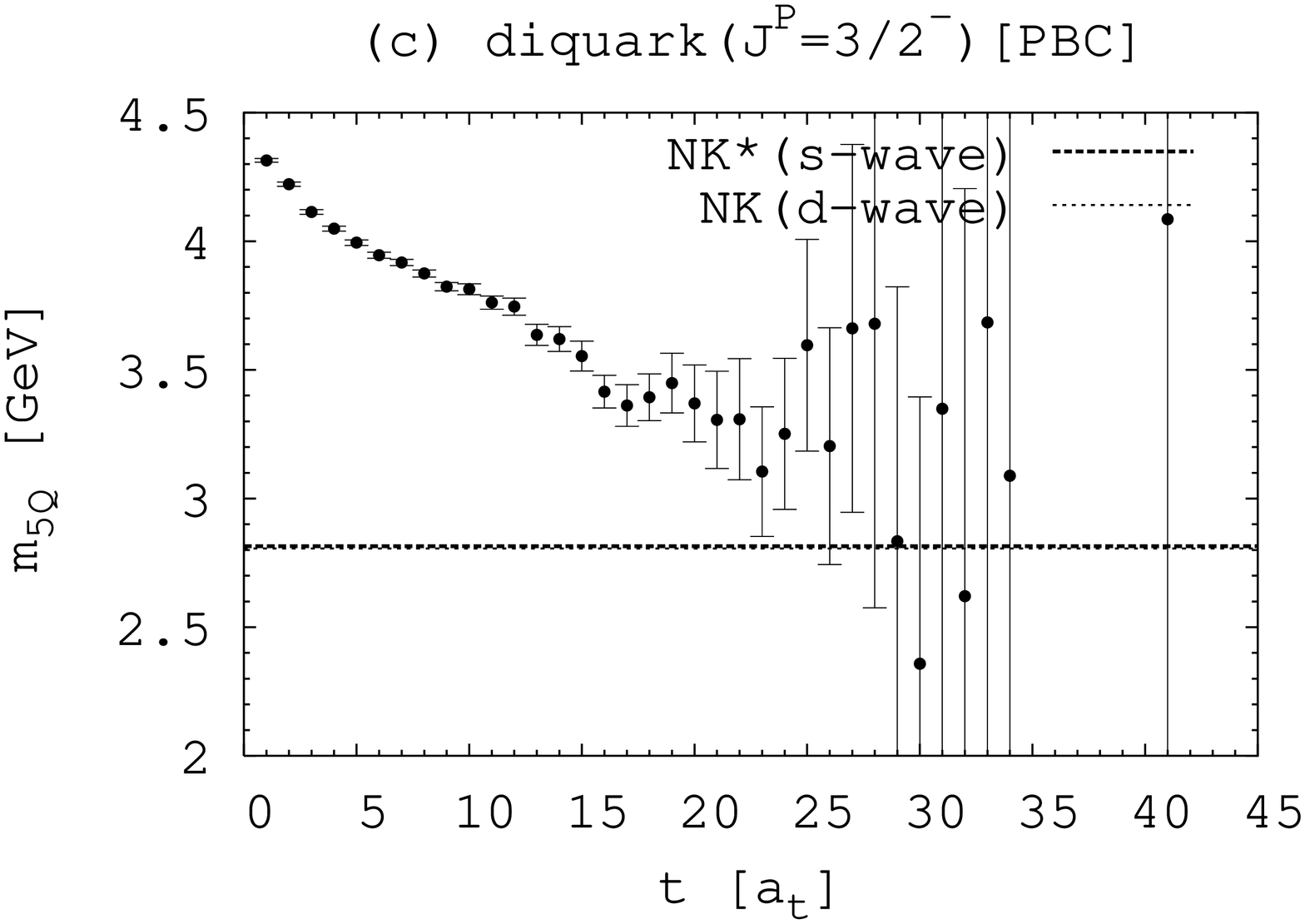}
\end{center}
\caption{The  5Q effective mass  plots in  $J^P=3/2^-$ channel  in the
standard  periodic BC(PBC)  for three  types of  interpolating fields,
i.e.,  (a) the  NK$^*$-type, (b)  the twisted  NK$^*$-type, and  (c) the
diquark-type.
\Eq{typical.set}   is   adopted   as   a  typical   set   of   hopping
parameters.  The statistical  error  is estimated  with the  jackknife
prescription.
The  dotted  lines  indicate  the  s-wave NK$^*$  and  the  d-wave  NK
threshold in the spatial lattice size $L\simeq 2.15$ fm.
Note that they accidentally coincide with each other.
The  solid lines  denote  the results  of  the single-exponential  fit
performed in each plateau region.
%
%
}
\label{fig.three.half.minus.pbc}
\end{figure}

We    consider    5Q   spectrum    in    $J^P=3/2^-$   channel.     In
\Fig{fig.three.half.minus.pbc},  we show the  effective mass  plots in
$J^P=3/2^-$  channel for  three  interpolating fields,  i.e., (a)  the
NK$^*$-type, (b) the twisted NK$^*$-type, (c) a diquark-type.
The  dotted  lines  indicate  the  s-wave NK$^*$  and  the  d-wave  NK
thresholds,    which    happen    to    coincide    accidentally    in
\Fig{fig.three.half.minus.pbc} for  the spatial lattice  size $L\simeq
2.15$ fm.

We define the effective mass as a function of $\tau$ by
\begin{equation}
  m_{\rm eff}(\tau)
  \equiv
  \log\left({ G^{(3/2)}(\tau) \over G^{(3/2)}(\tau+1) }\right),
\label{eq.effmass}
\end{equation}
where $G^{(3/2)}(\tau)$ denotes the temporal correlator.
At  sufficiently large  $\tau$,  the correlator  is  dominated by  the
lowest-lying state  with energy $m$  as $G^{(3/2)}(\tau) \sim  A e^{-m
\tau}$.  Then  \Eq{eq.effmass} gives a constant  as $m_{\rm eff}(\tau)
\simeq m$.
Thus a plateau in $m_{\rm eff}(\tau)$ indicates that the correlator is
saturated  by  a  single-state.   In  such cases,  we  can  perform  a
single-exponential fit in the plateau region.

\Fig{fig.three.half.minus.pbc}~(a) shows  the effective mass  plot for
the NK$^*$-type  interpolating field.  In  the region $0 \le  \tau \le
24$,  the  contamination  of  the higher  spectral  contributions  are
gradually  reduced, which  is indicated  by the  decreases  in $m_{\rm
eff}(\tau)$.
There is  a plateau in  the interval $25  \alt \tau \alt 35$,  where a
single-state is expected to  dominate the 5Q correlator.  Beyond $\tau
\sim  36$, the  statistical  error becomes  large.   In addition,  the
effect of the backward  propagation becomes gradually more significant
as $\tau\sim 48$ is approached.  Hence, we simply neglect the data for
$\tau \agt 36$,  and perform the single-exponential fit  in the region
$25  \le \tau  \le  35$.   We obtain  $m_{\rm 5Q}=2.90(2)$  GeV, which  is
denoted by the solid line.
One sees  that the 5Q states  appears above the s-wave  NK$^*$ and the
d-wave NK thresholds.

\Fig{fig.three.half.minus.pbc}~(b) shows  the effective mass  plot for
the twisted NK$^*$-type interpolating field.
There is a  plateau in the interval $24 \alt \tau  \alt 35$, where the
single-exponential  fit is performed  leading to  $m_{\rm 5Q}=2.89(1)$
GeV.
The  5Q state  is again  above  the s-wave  NK$^*$ and  the d-wave  NK
thresholds.

\Fig{fig.three.half.minus.pbc}~(c) shows  the effective mass  plot for
the diquark-type interpolating field.
We see that the statistical error is too large to identify the plateau
unambiguously.  Hence, we do not perform the fit.
Note that  this plot  is obtained by  using $N_{\rm  conf}=1000$ gauge
configurations.
A possible  reason for  such a large  noise is that  the interpolating
field  \Eq{diquark-type.SV}  does  not  survive  the  non-relativistic
limit due to the vector diquark.

\begin{figure}
\begin{center}
\includegraphics[height=0.48\textwidth,angle=-90]{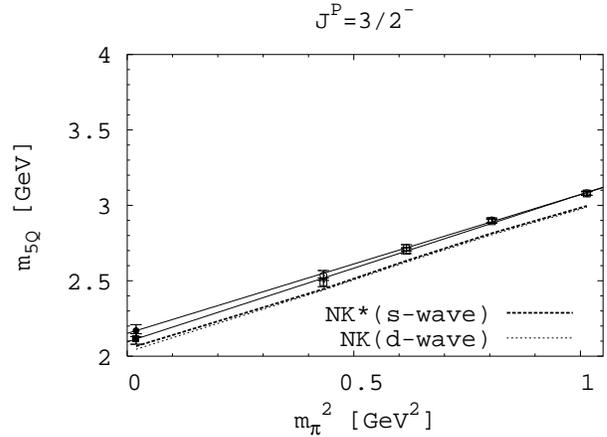}
\end{center}
\caption{$m_{\rm 5Q}$ in  $J^P=3/2^-$ channel against  $m_{\pi}^2$ for the
two interpolating  fields, i.e.,  (circle) the NK$^*$-type,  (box) the
twisted NK$^*$-type.
Open  symbols denote  the direct  lattice QCD  data, while  the closed
symbols  and the  solid  lines  represent the  results  of the  chiral
extrapolations to the physical quark mass region.
%
}
\label{fig.three.half.minus.chiral.pbc}
\end{figure}
\begin{table}
\begin{ruledtabular}
\begin{tabular}{lclllll}
$J^P$   & I.F. &0.1210 & 0.1220 & 0.1230 & 0.1240 & $\kappa_{\rm phys.}$ \\
\hline
$3/2^-$ & (a) & 3.08(1) & 2.90(2) & 2.72(2) & 2.54(3) & 2.17(4)
\\
$3/2^-$ & (b) & 3.08(1) & 2.89(1) & 2.70(2) & 2.49(3) & 2.11(4)
\\
$3/2^-$ & (c) & --       & --       & --       & --       & --
\\
$3/2^+$ &  (a) & 3.52(2) & 3.34(3) & 3.17(11) &3.00(5) & 2.64(7)
\\
$3/2^+$ &  (b) & 3.27(3) & 3.11(4) & 2.95(5) & 2.83(9) & 2.48(10)
\\
$3/2^+$ &  (c) & 3.34(2) & 3.16(2) & 2.98(3) & 2.78(5) & 2.42(6)
\end{tabular}
\end{ruledtabular}
\caption{
$m_{\rm  5Q}$ for each  value of  $\kappa$ in the physical unit GeV.
The  first column  labeled by  ``$J^P$''  indicates the  spin and  the
parity.
The  second column  labeled by  ``I.F.''  indicates  the interpolating
field used, i.e., (a) the  NK$^*$-type, (b) the twisted NK$^*$-type, (c)
the diquark-type.
$\kappa_{\rm phys.}\simeq 0.1261$ denotes  the value of $\kappa$ which
achieve $m_{\pi}\simeq 0.14$ GeV.
``--'' indicates  that the fitting is  not performed due  to the large
statistical error.  }
\label{table.chiral.extrapolation}
\end{table}
Now, we perform the chiral extrapolation. As mentioned before, we keep
$\kappa=0.1240$ fixed  for $s$-quark, and  vary $\kappa=0.1210-0.1240$
for $u$ and $d$ quarks.
\Fig{fig.three.half.minus.chiral.pbc}   shows   the   5Q   masses   in
$J^P=3/2^-$ channel against $m_{\pi}^2$.  Circles and boxes denote the
data  obtained from  the NK$^*$-type  and the  twisted  NK$^*$-type 5Q
correlators, respectively.
Note that they agree with each other within the statistical error.
%
%
%
The open symbols refer to the direct lattice QCD data.
Since these data  behave almost linearly in $m_{\pi}^2$,  we adopt the
linear chiral  extrapolation in $m_{\pi}^2$ to obtain  $m_{\rm 5Q}$ in
the physical quark mass region.   Note that the ordinary non-PS mesons
and baryons show similar linearity in $m_{\pi}^2$ \cite{nemoto}.
The closed symbols denote the results of the chiral extrapolation.
%
%
We see that  all the 5Q states appear above the  s-wave NK$^*$ and the
d-wave NK thresholds.
%
As a  result of  the chiral extrapolation,  we obtain only  massive 5Q
states as  $m_{\rm 5Q}=2.17(4), 2.11(4)$ GeV from the  NK$^*$-type and the
twisted NK$^*$-type correlators, respectively,
which is too  heavy to be identified with  the experimentally observed
$\Theta^+(1540)$.
Numerical values  of $m_{\rm 5Q}$ at each hopping  parameter together with
their    chirally    extrapolated    values    are    summarized    in
\Table{table.chiral.extrapolation}.
%
%
To obtain a low-lying state at $m_{\rm 5Q}\simeq 1540$ MeV, a 5Q state
should appear below these thresholds  at least in the light quark mass
region. In this case, a significantly large chiral effect is required.
Of course,  this point  can be in  principle clarified by  an explicit
lattice QCD calculation with chiral fermions.
%
%


\subsection{$J^P=3/2^+$ 5Q spectrum in PBC}
We consider 5Q spectrum in $J^P=3/2^+$ channel.
$J^P=3/2^+$ is  an interesting quantum  number from the view  point of
the  diquark  picture  of  Jaffe  and  Wilczek\cite{jaffe}.   In  this
picture,  the pair  of diquarks  has  angular momentum  one, which  is
combined with the  spin 1/2 of $\bar{s}$ quark.   Hence, there are two
possibilities as  $J^P=1/2^+$ and  $3/2^+$, i.e., the  diquark picture
can support $J^P=3/2^+$ possibility as well.
Its mass splits  from the $J^P=1/2^+$ state depending  on a particular
form  of the  LS-interaction.
%
%
If it is massive,  it is expected to have a large  decay width.  If it
is light  enough,
its exotic structure  may work to implement the  narrow decay width as
in $J^P=1/2^+$ case.

\begin{figure}
\begin{center}
\includegraphics[height=0.48\textwidth,angle=-90]{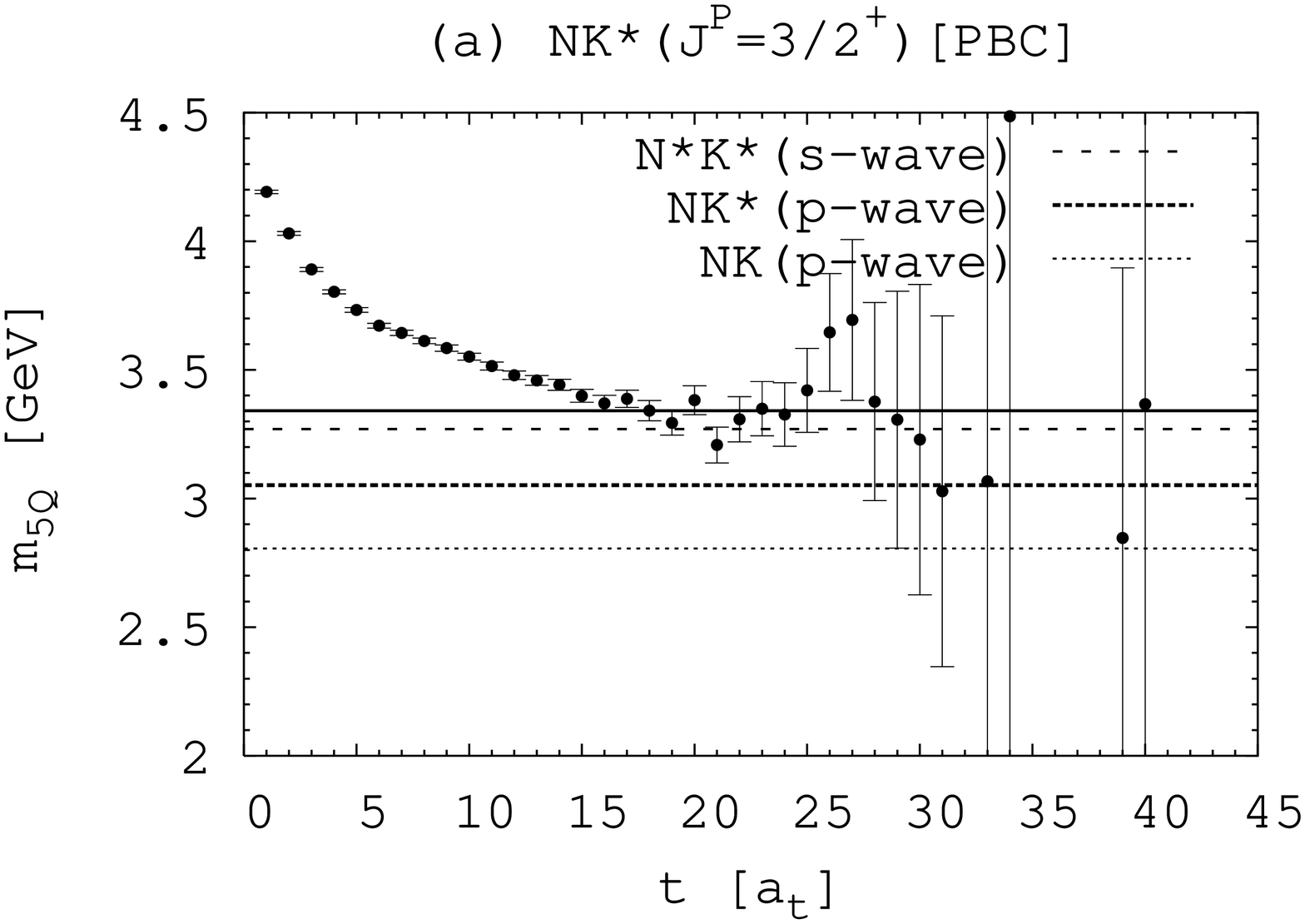}  
\\
\includegraphics[height=0.48\textwidth,angle=-90]{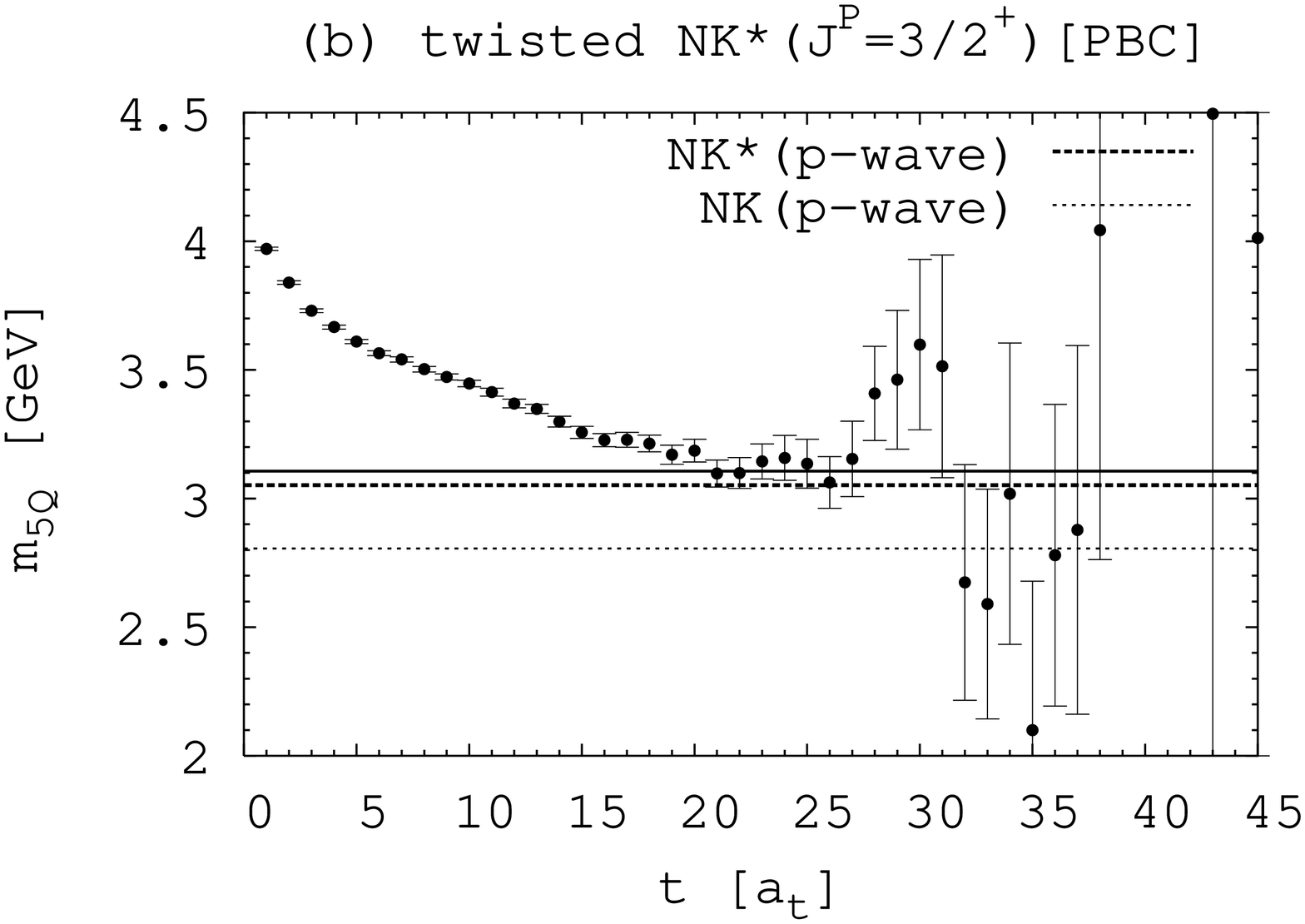}  
\\
\includegraphics[height=0.48\textwidth,angle=-90]{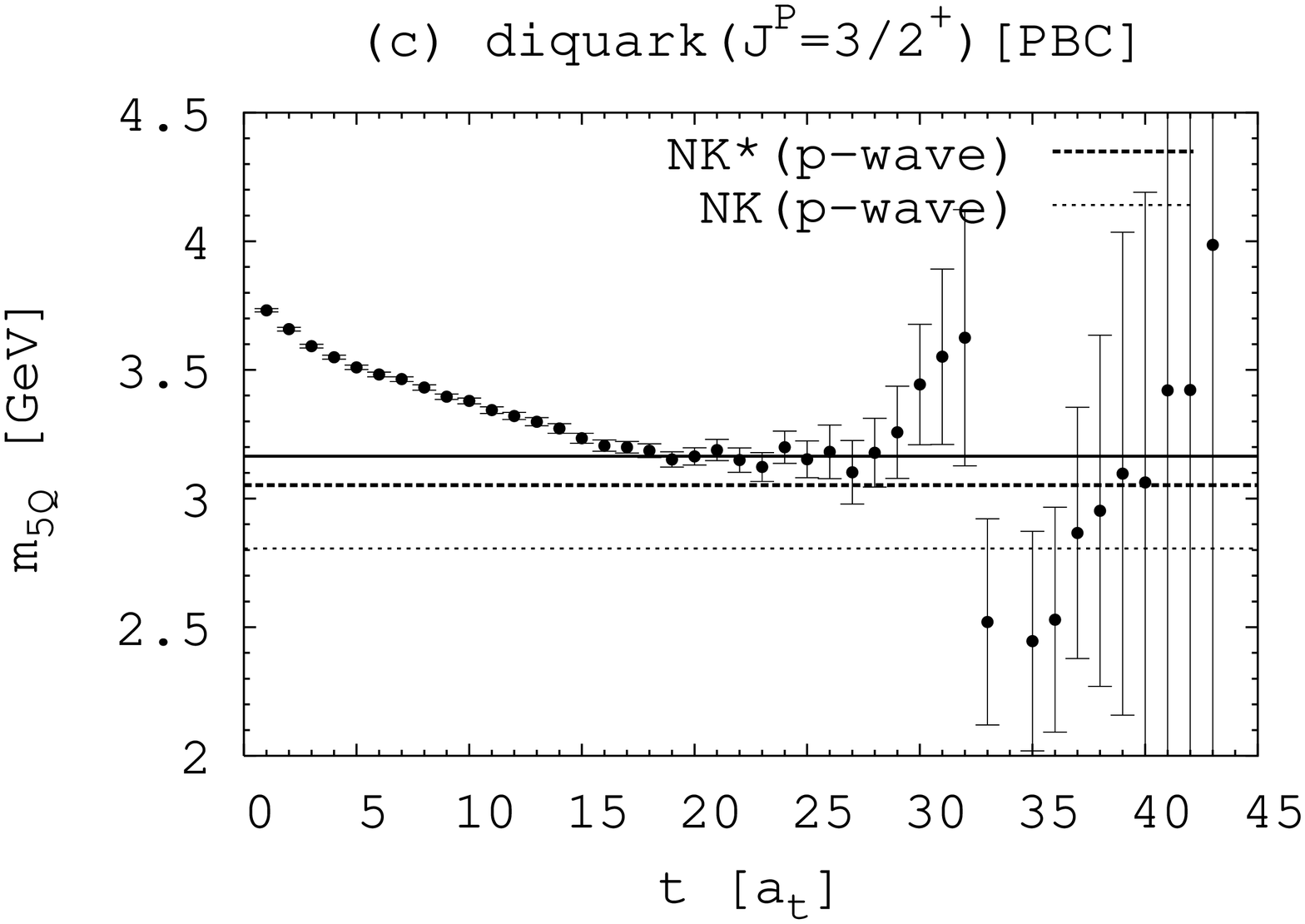}
\end{center}
\caption{The 5Q effective mass plots in $J^P=3/2^+$ channel in PBC for
three types  of interpolating fields,  i.e., (a) the  NK$^*$-type, (b)
the twisted NK$^*$-type, and (c) the diquark-type.
\Eq{typical.set}   is   adopted   as   a  typical   set   of   hopping
parameters.
The dotted lines indicate the s-wave N$^*$K$^*$, the p-wave NK$^*$ and
the p-wave  NK thresholds in  the spatial lattice size  $L\simeq 2.15$
fm.
The  solid lines  denote  the results  of  the single-exponential  fit
performed in each plateau region.
}
\label{fig.three.half.plus.pbc}
\end{figure}

In \Fig{fig.three.half.plus.pbc}, we show  the 5Q effective mass plots
in PBC  employing three types  of interpolating fields, i.e.,  (a) the
NK$^*$-type, (b) the twisted NK$^*$-type, (c) the diquark-type.
The dotted lines indicate the s-wave N$^*$K$^*$, the p-wave NK$^*$ and
the p-wave NK  thresholds in the spatial lattice  of the size $L\simeq
2.15$ fm, respectively.
\begin{figure}
\begin{center}
\includegraphics[height=0.48\textwidth,angle=-90]{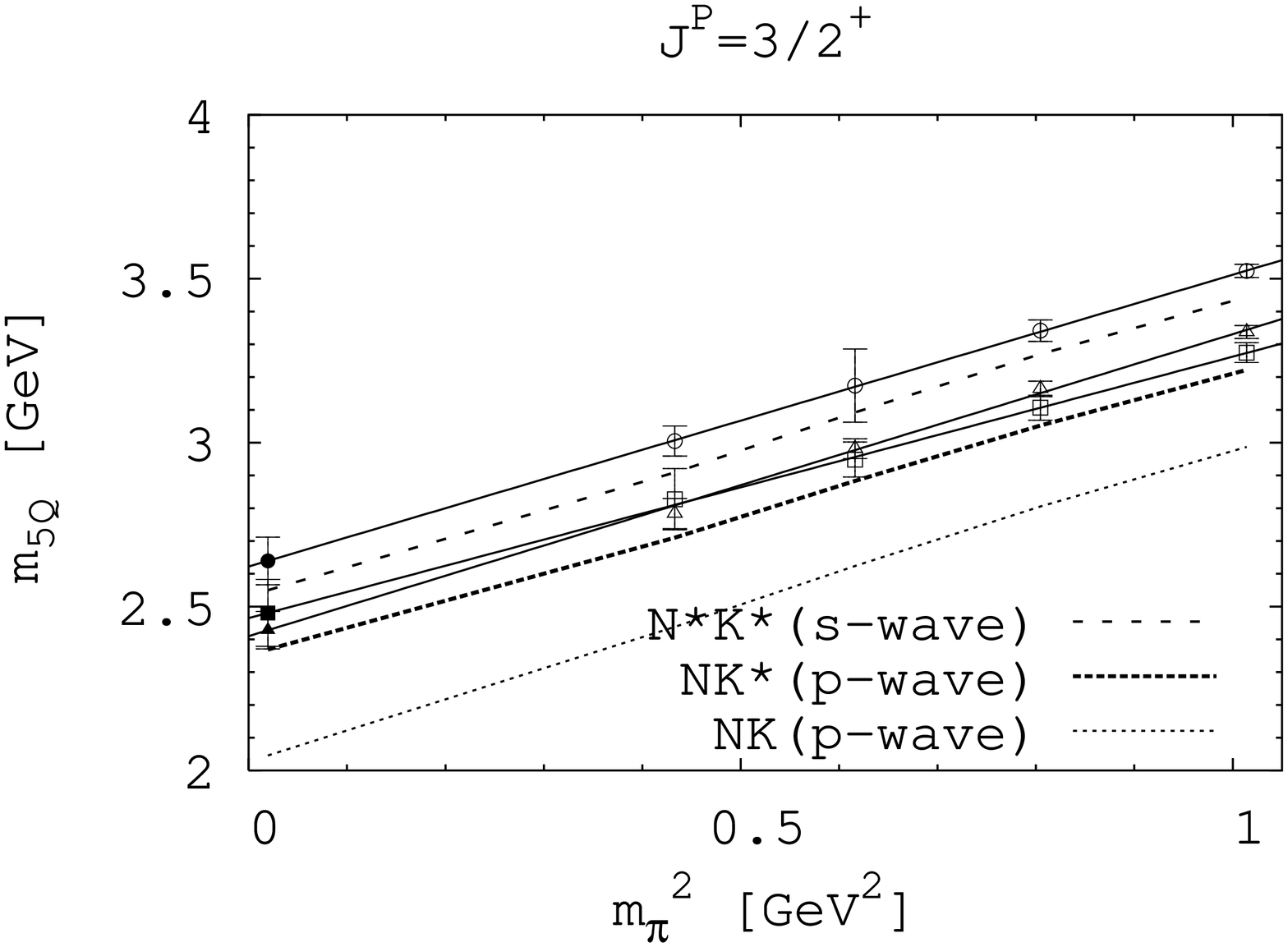}
\end{center}
\caption{$m_{\rm 5Q}$ in  $J^P=3/2^+$ channel against  $m_{\pi}^2$ for the
three interpolating fields, i.e.,  (circle) the NK$^*$-type, (box) the
twisted      NK$^*$-type,     and     (triangle)      a     diquark-type
[\Eq{diquark-type.SV}].
Open  symbols denote  the direct  lattice QCD  data, while  the closed
symbols  and the  solid  lines  represent the  results  of the  chiral
extrapolations to the physical quark mass region.
%
%
}
\label{fig.three.half.plus.chiral.pbc}
\end{figure}

\Fig{fig.three.half.plus.pbc}~(a)  shows the  5Q  effective mass  plot
employing the  NK$^*$-type interpolating  field. In the  region, $0\le
\tau  \alt 17$,  the contaminations  of higher  spectral contributions
become gradually reduced.
There is a flat region $18 \alt \tau \le 30$, which is still afflicted
with slightly large statistical errors.
The single-exponential fit in this region gives $m_{\rm 5Q}=3.34(3)$ GeV.
Note  that this  value  agrees with  the  s-wave N$^*$K$^*$  threshold
$E_{\rm th}\simeq 3.27$  GeV.  (See \Table{table.mass} for $m_{N^*}$.)

\Fig{fig.three.half.plus.pbc}~(b)  shows the  5Q  effective mass  plot
corresponding to the twisted  NK$^*$-type interpolating field. We have
a rather  stable plateau in the  interval $21 \le \tau  \le 27$, where
the   single-exponential   fit  is   performed.   We  obtain   $m_{\rm
5Q}=3.11(4)$ GeV.  The result is denoted by the solid line.

\Fig{fig.three.half.plus.pbc}~(c) shows the 5Q effective mass plot for
the  diquark-type  interpolating field.   We  find  a  plateau in  the
interval $19  \le \tau  \le 29$, where  the single-exponential  fit is
performed.  We  obtain $m_{\rm 5Q}=3.16(2)$ GeV,  which is denoted  by the
solid line.

Now,  we  perform  the  chiral  extrapolation.
In  \Fig{fig.three.half.plus.chiral.pbc}, $m_{\rm 5Q}$ is  plotted against
$m_{\pi}^2$.  Circles,  boxes and  triangles denote the  data obtained
from the  NK$^*$-type, the twisted  NK$^*$-type and the  diquark-type 5Q
correlators, respectively.
Note that the latter two  agree with each other within the statistical
errors.
%
%

As  a   result  of  the   chiral  extrapolation,  we   obtain  $m_{\rm
5Q}=2.64(7)$ GeV from the NK$^*$-type correlator,
$m_{\rm 5Q}=2.48(10)$ GeV from  the twisted NK$^*$-type correlator,
and $m_{\rm 5Q}=2.42(6)$ GeV from the diquark-type correlator.
Numerical values  of $m_{\rm 5Q}$ in  $J^P=3/2^+$ channel at  each hopping
parameter together  with their  chirally extrapolated values  are also
summarized in \Table{table.chiral.extrapolation}.
The  two  data  from   the  twisted  NK$^*$-type  and  the  diquark-type
correlators  are considered to  be almost  consistent with  the p-wave
NK$^*$ threshold,
while the data from the  NK$^*$-type correlator seems to correspond to
a  more massive  state,  which is  likely  to be  consistent with  the
N$^*$K$^*$(s-wave) threshold.
%
We see  again that all  of our data  of $m_{\rm 5Q}$ appear  above the
NK$^*$  threshold(p-wave),  which is  located  above the  artificially
raised  NK threshold(p-wave)  due  to the  finiteness  of the  spatial
lattice as $L\simeq 2.15$ fm.  As a result, we are left only with such
massive 5Q states.

Now, several comments are in order.
(1) \Ref{lasscock32} reported the existence of a low-lying 5Q state in
$J^P=3/2^+$ channel  using NK$^*$-type interpolating  field.  However,
we have not observed such a low-lying 5Q state in our calculation.
%
%
(2) Recall that,  except for a  single calculation\cite{chiu}, lattice
QCD   calculations   indicate   that   $J^P=1/2^+$  state   is   heavy
\cite{scikor12,   sasaki,  kentacky,   ishii12,   rabbit,  lasscock12,
  alexandrou12, csikor122,  holland}, for instance  $m_{\rm 5Q} \simeq
2.25$ GeV in \Ref{ishii12}.
From  the viewpoint of  the diquark  picture, it  could be  natural to
obtain such massive 5Q states in $J^P=3/2^+$ channel.
If there  were a low-lying 5Q  state in $J^P=3/2^+$  channel, then the
diquark picture could suggest also a low-lying 5Q state in $J^P=1/2^+$
channel nearby.

\section{Analysis with HBC}
\label{HBC.analysis}
\begin{figure}
\begin{center}
\includegraphics[height=0.48\textwidth,angle=-90]{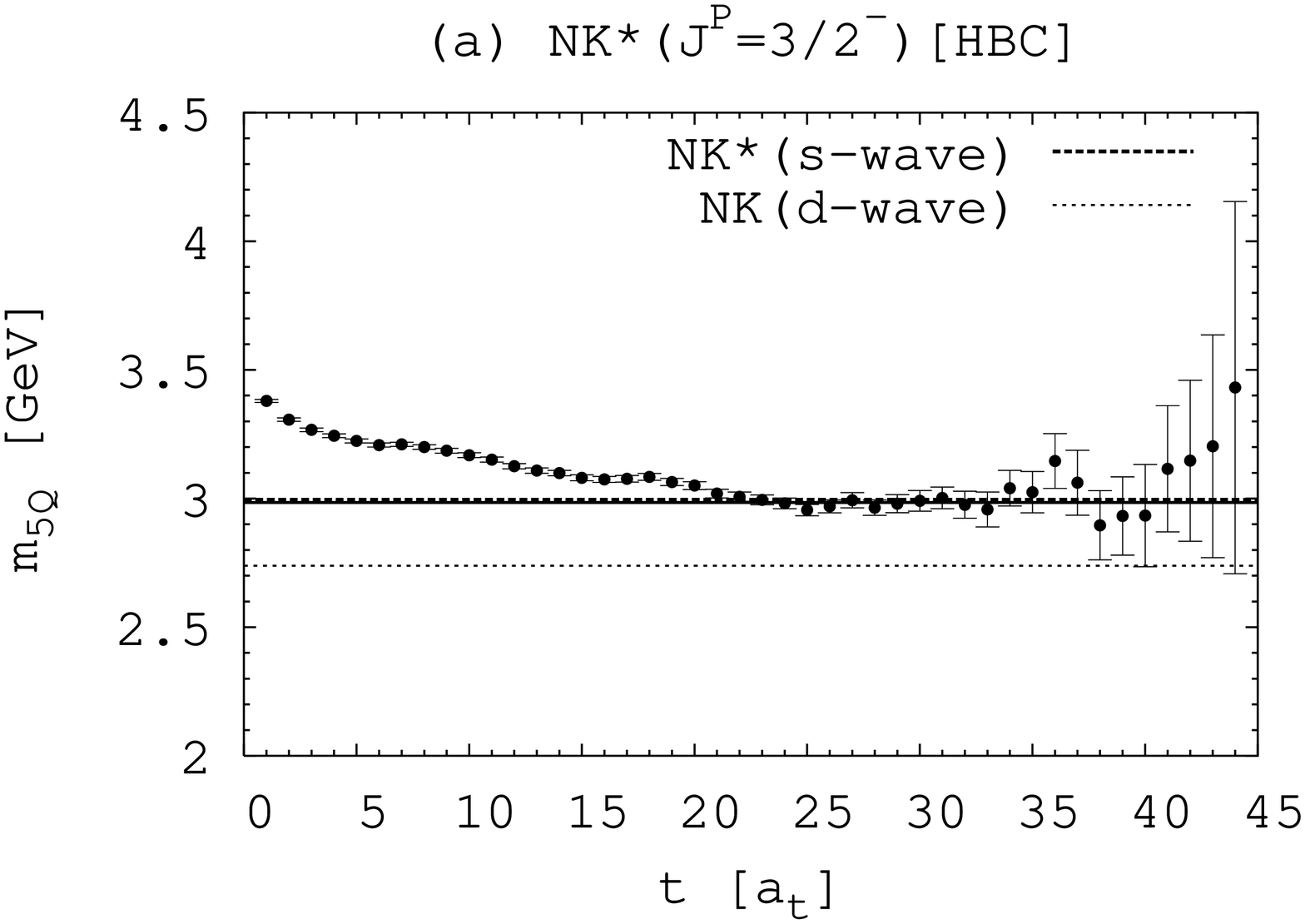}  
\\
\includegraphics[height=0.48\textwidth,angle=-90]{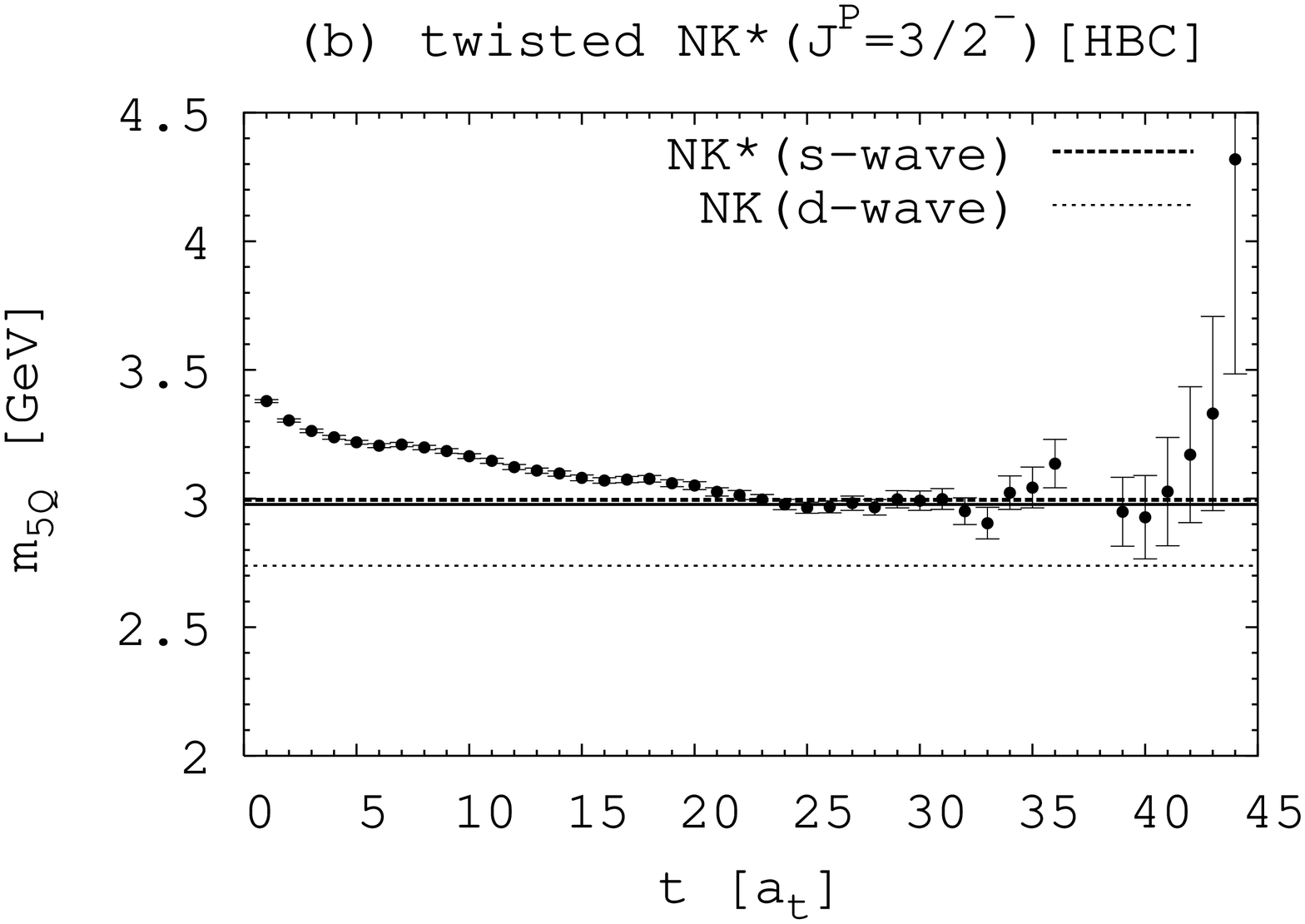}  
\\
\includegraphics[height=0.48\textwidth,angle=-90]{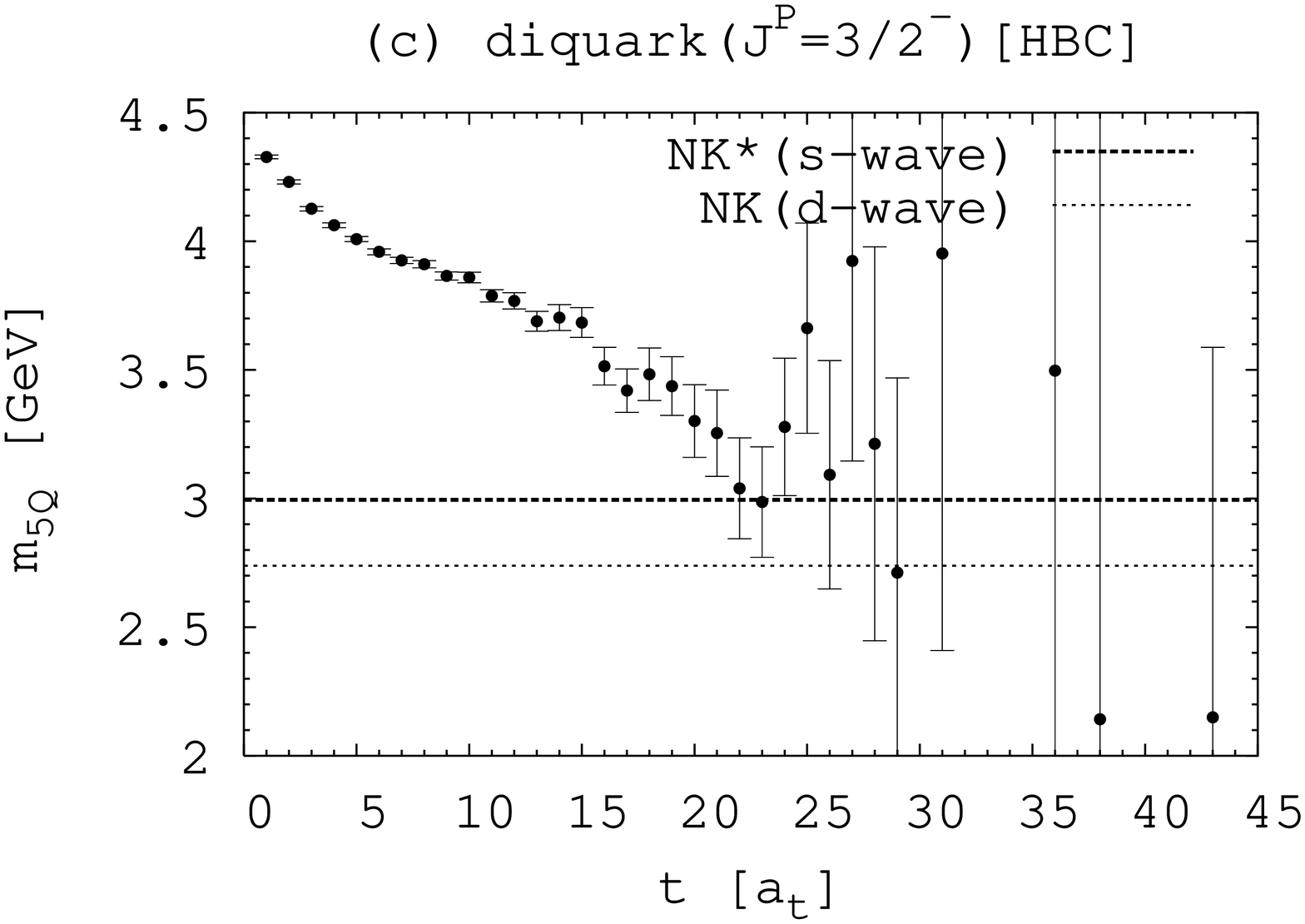}
\end{center}
\caption{The 5Q effective mass plots in $J^P=3/2^-$ channel in HBC for
three types  of interpolating fields,  i.e., (a) the  NK$^*$-type, (b)
the twisted NK$^*$-type, and (c) the diquark-type.
The meanings of  the dotted lines and the solid lines  are the same as
in \Fig{fig.three.half.minus.pbc}.
%
%
}
\label{fig.three.half.minus.hbc}
\vspace{6ex}
\end{figure}
In the  previous section,  we have only  massive 5Q states,  which are
obtained by using the linear chiral extrapolation in $m_{\pi}^2$.
However, the chiral  behavior may deviate from a  simple linear one in
the light quark mass region, which could lead to somewhat less massive
states.
Considering  this, we  think  it of  worth  at this  stage to  analyze
whether our 5Q states are compact 5Q resonances or not.
This  is done  by  switching the  spatial  periodic BC  to the  hybrid
BC(HBC) introduced in \Sect{general.formalisms}.

\subsection{$J^P=3/2^-$ 5Q spectrum in HBC}
\Fig{fig.three.half.minus.hbc}  shows the 5Q  effective mass  plots in
HBC employing the  three types of interpolating fields,  i.e., (a) the
NK$^*$-type, (b) the twisted NK$^*$-type, and (c) the diquark type.
These  figures  should be  compared  with  their  PBC counterparts  in
\Fig{fig.three.half.minus.pbc}.
The  dotted  lines  denote  the   s-wave  NK$^*$  and  the  d-wave  NK
thresholds.
%
For the typical set of hopping parameters, i.e., \Eq{typical.set}, the
s-wave NK$^*$ threshold(the  thick dotted line) is raised  up by $\sim
180$ MeV, and the d-wave NK threshold(the thin dotted line) is lowered
down  by $\sim 70$  MeV due  to HBC  in the  finite spatial  extent as
$L\simeq 2.15$ fm.  (See \Table{thresholds}.)

\Fig{fig.three.half.minus.hbc}~(a)  shows the  5Q effective  mass plot
for the NK$^*$-type interpolating field in HBC.
We find  a plateau  in the interval  $23 \le  \tau \le 35$,  where the
single-exponential fit is performed  leading to $m_{\rm 5Q}= 2.98(1)$ GeV,
which is denoted by the solid line.
We see that $m_{\rm 5Q}$ is raised up by 80 MeV due to HBC.
%
The  value  of $m_{\rm  5Q}$  is  consistent  with the  s-wave  NK$^*$
threshold within the statistical error.
Therefore, we regard this state as an NK$^*$ scattering state.

\Fig{fig.three.half.minus.hbc}~(b)  shows the  5Q effective  mass plot
for the twisted NK$^*$-type interpolating field.
We find  a plateau  in the interval  $24 \le  \tau \le 35$,  where the
single exponential fit is performed leading to $m_{\rm 5Q} = 2.98(1)$ GeV,
which is denoted by the solid line.
The situation is similar to the NK$^*$-interpolating field case.
We see that $m_{\rm 5Q}$ is raised up  by 90 MeV due to HBC.
%
%
Since the value is consistent  with the s-wave NK$^*$ threshold within
the statistical error, we regard it as an NK$^*$ scattering state.

\Fig{fig.three.half.minus.hbc}~(c) shows the 5Q effective mass plot
for the diquark-type interpolating field.
We  see that  it is  afflicted with  considerable size  of statistical
errors as before, due to which the best-fit is not performed.
%

%
In this way,  all of our 5Q states in $J^P=3/2^-$  channel turn out to
be NK$^*$  scattering states.  More  precisely, we do not  observe any
compact 5Q  resonance states in  $J^P=3/2^-$ channel below  the raised
s-wave NK$^*$ threshold, i.e., in the following region:
\begin{equation}
  E
  \alt 
  \sqrt{m_N^2 + \vec p_{\rm min}^2}
  +
  \sqrt{m_{K^*}^2 + \vec p_{\rm min}^2},
\end{equation}
with $|\vec p_{\rm min}|\simeq 499$ MeV.

\subsection{$J^P=3/2^+$ 5Q spectrum in HBC}
\begin{figure}[htb]
\begin{center}
\includegraphics[height=0.48\textwidth,angle=-90]{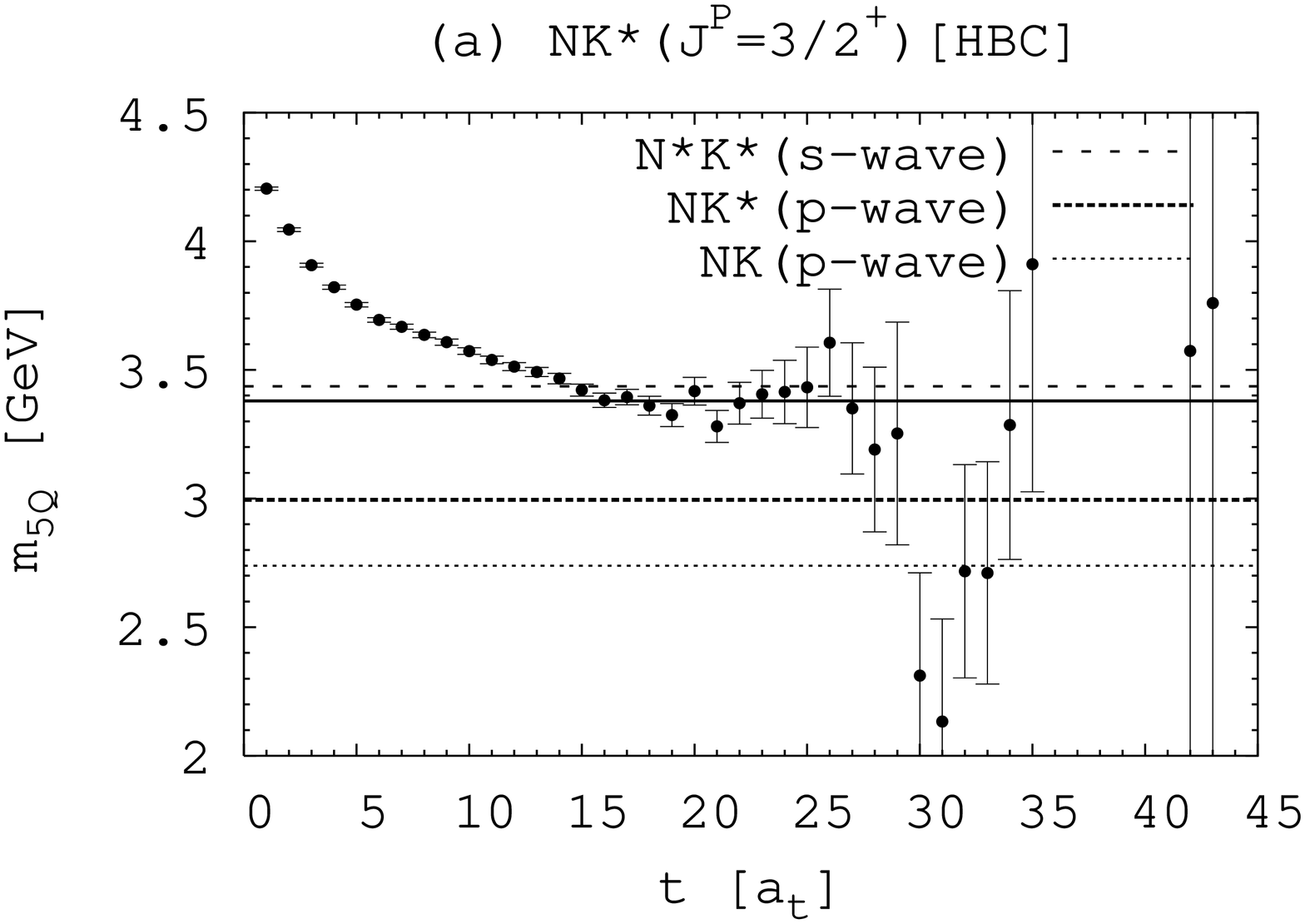}  
\\
\includegraphics[height=0.48\textwidth,angle=-90]{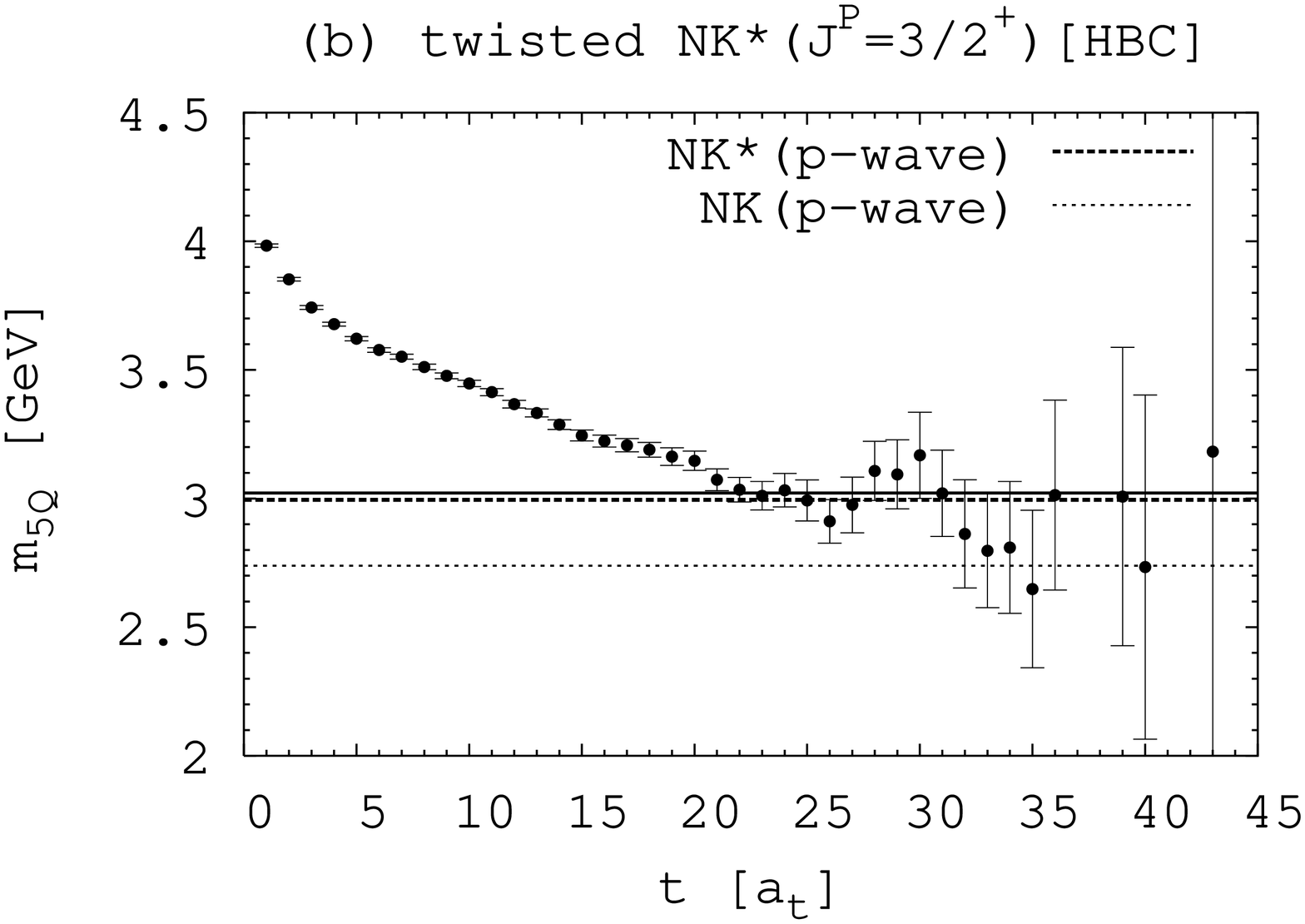}  
\\
\includegraphics[height=0.48\textwidth,angle=-90]{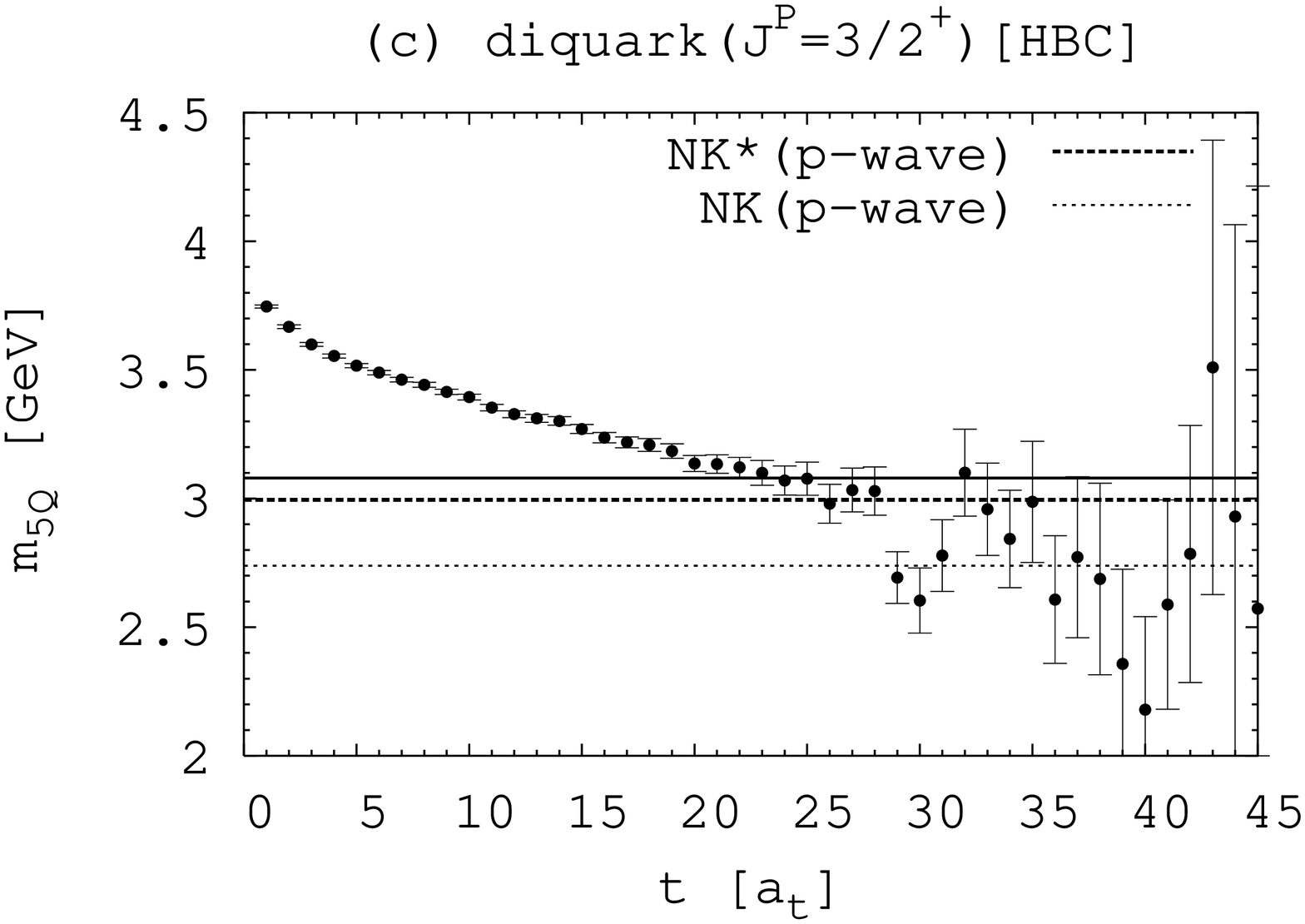}
\end{center}
\caption{5Q  effective mass plots  in $J^P=3/2^+$  channel in  HBC for
three types of interpolating fields (a) the NK$^*$-type, (b) the twisted
NK$^*$-type, and (c) the diquark-type.
The meanings  of the  dotted and the  solid lines  are the same  as in
\Fig{fig.three.half.plus.pbc}.
}
\label{fig.three.half.plus.hbc}
\end{figure}
\Fig{fig.three.half.plus.hbc} shows the 5Q effective mass plots in HBC
employing  the three  types  of interpolating  fields,  i.e., (a)  the
NK$^*$-type,  (b) the  twisted  NK$^*$-type, and  (c) the  diquark-type.
These  figures  should be  compared  with  their  PBC counterparts  in
\Fig{fig.three.half.plus.pbc}.  The  meanings of  the  dotted and  the
solid lines are the same as in \Fig{fig.three.half.plus.pbc}.

HBC may  not be useful in  $J^P=3/2^+$ channel, since  it induces only
minor changes in the two-particle spectra.
For the typical set of hopping parameters, i.e., \Eq{typical.set}, the
p-wave NK$^*$ threshold is lowered down only by $\sim 60$ MeV, and the
p-wave NK threshold is lowered down only by $\sim 70$ MeV.
We see  that these shifts are  rather small. This is  because they are
induced  by the changes  in the  minimum non-vanishing  momentum, i.e.,
$|\vec p_{\rm min}|=2\pi/L\simeq 576$ MeV to $\sqrt{3}\pi/L\simeq 499$
MeV as mentioned before.
In  $J^P=3/2^+$ channel, N$^*$K$^*$(s-wave)  threshold shows  the most
drastic change, i.e., the upper  shift by 170 MeV, which however plays
a less significant role, since its location is at rather high energy.

\Fig{fig.three.half.plus.hbc}~(a)  shows the  5Q  effective mass  plot
employing the NK$^*$-type interpolating field.
There is a flat region $16 \le \tau \alt 25$, which is still afflicted
with  slightly large  statistical errors.
%
The   single-exponential  fit   in  this   region  leads   to  $m_{\rm
5Q}=3.38(2)$ GeV,  which is  denoted by the  solid line.  We  see that
$m_{\rm 5Q}$ is raised up by 40 MeV.
Although the  shift of 40 MeV  is rather small, $m_{\rm  5Q}$ is again
almost  consistent  with the  s-wave  N$^*$K$^*$  threshold.
Considering  its rather  large  statistical error,  this  5Q state  is
likely to be an s-wave N$^*$K$^*$ scattering state.
To draw  a more  solid conclusion  on this state,  it is  necessary to
improve the statistics further more.
%

\Fig{fig.three.half.plus.hbc}~(b)  shows the  5Q  effective mass  plot
employing the twisted NK$^*$-type interpolating field.
There is  a plateau  in the interval  $23 \le  \tau \le 31$,  where we
perform the single-exponential fit. The result $m_{\rm 5Q}=3.02(3)$ GeV is
denoted by the solid line.  We see that $m_{\rm 5Q}$ is lowered down by 90
MeV,  which is  considered  to be  consistent  with the  shift of  the
NK$^*$(p-wave)  threshold.
Therefore, this state is likely to be an NK$^*$(p-wave) scattering state.

\Fig{fig.three.half.plus.hbc}~(c)  shows the  5Q  effective mass  plot
employing the diquark-type interpolating field.
Although  the  data is  slightly  noisy, there  is  a  plateau in  the
interval  $23 \le  \tau \le  28$.   A single-exponential  fit in  this
plateau region leads to $m_{\rm 5Q}=3.08(4)$  GeV, which is denoted by the
solid line.
$m_{\rm 5Q}$ is lowered down by 80 MeV due to HBC.
%
The situation is  similar to \Fig{fig.three.half.plus.hbc}~(b).
This state is likely to be an NK$^*$(p-wave) scattering state.


In  this  way,  all  of  our   5Q  states  are  likely  to  be  either
N$^*$K$^*$(s-wave)  or NK$^*$(p-wave)  states rather  than  compact 5Q
resonance states.
Of course, because  HBC induces only minor changes  in the 5Q spectrum
in $J^P=3/2^+$ channel, and  also because 5Q correlators still involve
considerable size  of statistical error, more  statistics is desirable
to draw a more solid conclusion on the real nature of these 5Q states.
Here, we  can at  least state  that these 5Q  states are  all massive,
which locate above NK$^*$(p-wave) threshold.


\section{Summary and conclusion}
\label{summary}
We have  studied $J^P=3/2^\pm$ penta-quark(5Q)  baryons in anisotropic
lattice QCD at  the quenched level
with  a  large  number   of  gauge  field  configurations  as  $N_{\rm
conf}=1000$ for high precision measurements.
We emphasize that the spin  of $\Theta^+(1540)$ has not yet determined
experimentally, and  that the $J^P=3/2^-$ assignment  provides us with
one of the possible solutions to  the puzzle of the narrow decay width
of $\Theta^+(1540)$ \cite{hosaka32}.
We have employed  the standard Wilson gauge action  on the anisotropic
lattice of  the size $12^3\times 96$ with  the renormalized anisotropy
$a_{\rm s}/a_{\rm  t} = 4$  at $\beta=5.75$, which  leads to
$a_{\rm s} \simeq 0.18$ fm and $a_{\rm t} \simeq 0.045$ fm.
We have found that correlators of 5Q baryons in $J^P=3/2^\pm$ channels
are rather noisy.  Hence,  the large statistics as $N_{\rm conf}=1000$
has played a key role to get a solid result in our calculation.
%
For the  quark part, we have employed  $O(a)$-improved Wilson (clover)
action   with    four   values   of   the    hopping   parameters   as
$\kappa=0.1210(0.0010)0.1240$,  which  roughly  cover the  quark  mass
region  as $m_s \alt  m \alt  2m_s$.  To  avoid the  contaminations of
higher spectral contributions, we have employed the spatially extended
source in the 5Q correlators.

We have examined  several types of the 5Q  interpolating fields as (a)
the  NK$^*$-type,   (b)  the  (color-)twisted   NK$^*$-type,  (c)  the
diquark-type.
%
%
%
In $J^P=3/2^-$ channel, there are plateaus in the effective mass plots
for  the NK$^*$-type  and the  twisted NK$^*$-type  interpolating field,
whereas no  plateau has been  identified in that for  the diquark-type
interpolating field due to the significantly large statistical error.
The former two give almost identical results.
We  have employed  the  linear chiral  extrapolations in  $m_{\pi}^2$,
which  have  lead  to  $m_{\rm 5Q}\simeq  2.17$ and  $2.11$  GeV  for  the
NK$^*$-type and the twisted NK$^*$-type 5Q correlators, respectively.
%
%
In $J^P=3/2^+$ channel,  we have recognized plateaus in  all the three
effective  mass  plots.   However,  the plateau  for  the  NK$^*$-type
interpolating field  is located at  a somewhat higher energy  than the
other two.
The chiral extrapolations have lead to $m_{\rm 5Q}\simeq 2.64$ GeV for
the  NK$^*$-type  correlator,  $m_{\rm  5Q}\simeq 2.48$  GeV  for  the
twisted NK$^*$-type  correlator, and  $m_{\rm 5Q}\simeq 2.42$  GeV for
the diquark-type correlator.
%
%
In this way,  our data have not supported low-lying  5Q states in both
$J^P=3/2^{\pm}$ channels.
All the 5Q  states have been observed to appear  above the d/p-wave NK
threshold, which is artificially raised up by a few hundred MeV due to
the finiteness of the spatial  lattice as $L\simeq 2.15$ fm.
Note that, to  obtain low-lying 5Q states in  $J^P=3/2^\pm$ channel, a
5Q  state should  appear below  the raised  NK  threshold(p/d-wave) at
least in the light quark mass region.

In  order  to clarify  whether  the  observed  states are  compact  5Q
resonances or not, we have  performed an analysis with hybrid boundary
condition(HBC), which was recently proposed by \Ref{ishii12}.
In $J^P=3/2^-$ channel, our 5Q  states observed in the NK$^*$-type and
the  twisted NK$^*$-type  correlators  have turned  out  to be  s-wave
NK$^*$ scattering states.
In  $J^P=3/2^+$   channel,  for   the  twisted  NK$^*$-type   and  the
diquark-type correlators, the observed 5Q states are most likely to be
NK$^*$(p-wave) scattering states.
For the other one, i.e., the NK$^*$-type interpolating field, although
more statistics  is needed to draw  a definite conclusion,  it is most
likely to be an s-wave N$^*$K$^*$ scattering state.
Note that,  since HBC  does not affected  the two-particle  spectra so
much in $J^P=3/2^+$  channel, it is not easy  to elucidate the natures
of the 5Q states only with HBC.
At any rate, whatever the real  nature of these 5Q states may be, they
result in  a considerably  massive states in  the physical  quark mass
region,  which  cannot   be  identified  as  $\Theta^+(1540)$  without
involving a significantly large chiral contribution.
%
%

In this way,  we have not obtained any  relevant signals for low-lying
compact 5Q resonance states in $J^P=3/2^\pm$ channel below 2.1 GeV
in this paper, 
although the  $J^P=3/2^-$ possibility
provides us  with one of the  possible solutions to the  puzzle of the
narrow decay width of $\Theta^+(1540)$.
To get more solid conclusion on the pentaquark, 
it is important to perform the systematic studies of the 5Q states 
with the various  quantum numbers in lattice QCD 
with more sophisticated conditions. For instance, it is desired to use 
(1) unquenched full lattice QCD, 
(2) finer and larger volume lattice,  
(3) chiral fermion with small mass,
(4) more sophisticated interpolating field corresponding to the 
diquark picture and so on.
In any case,  the mysterious exotic hadron of  the pentaquark would be
much clarified  in future  studies of  lattice QCD as  well as  in the
future experiments.

\begin{acknowledgments}
We thank  A.~Hosaka, J.~Sugiyama, T.~Shinozaki  for useful information
and discussions.
M.~O and H.~S  are supported in part by  Grant for Scientific Research
((B)  No.   15340072 and  (C)  No.   16540236)  from the  Ministry  of
Education, Culture, Sports, Science and Technology, Japan.
T.~D. is supported by Special Postdoctoral Research Program of RIKEN.
Y.~N. is supported by 21st Century COE Program of Nagoya University.
%
%
%
The  lattice  QCD Monte  Carlo  calculations  have  been performed  on
NEC-SX5 at Osaka University.
\end{acknowledgments}


\appendix
\section{Spectral representation}
\label{spectral.representation}
Considering the importance of  the parity determination of $\Theta^+$,
we  present  a brief  derivation  of  the  spectral representation  of
Rarita-Schwinger    correlators,    i.e.,   \Eq{correlator.1},    with
\Eq{correlator.2}.
In this  section, gamma matrices  are represented in  Minkowskian form
(See \Ref{itzykson}).
To  avoid  unnecessary complexities,  we  derive only  $J^P=3/2^{\pm}$
parts.   The  $J^P=1/2^{\pm}$  parts  can  be  obtained  as  a  slight
modification,  on which  we will  make  a comment  at the  end of  the
section.

We   first  consider   the  coupling   of  our   interpolating  fields
$\psi_{\mu}$ to $J^P=3/2^{\pm}$ (anti-)baryon states.
Due    to    \Eq{parity},     our    interpolating    fields,    i.e.,
Eqs.~(\ref{nkstar}),(\ref{n8k8star})  and (\ref{diquark-type.SV}) have
the negative intrinsic parity.   Hence, their couplings to $J^P=3/2^-$
(anti-)baryons are parameterized in the following way:
\begin{eqnarray}
  \langle 0|
  \psi_{\mu}(0)
  |B_{3/2^-}(k,\alpha)\rangle
  &=&
  \lambda_{3/2^-} u_{\mu}(m_{3/2^-}; k,\alpha)
  \label{coupling.1}
  \\\nonumber
  \langle 0|
  \bar \psi_{\mu}(0)
  |\bar{B}_{3/2^-}(k,\alpha)\rangle
  &=&
  \lambda_{3/2^-}^*
  \bar v_{\mu}(m_{3/2^-}; k,\alpha),
\end{eqnarray}
where         $|B_{3/2^-}(k,\alpha)\rangle$         and         $|\bar
B_{3/2^-}(k,\alpha)\rangle$  denote  $J^P=3/2^-$ (anti-)baryon  states
with momentum $k$, helicity $\alpha$, and mass $m_{3/2^+}$.
$u_{\mu}(m;   k,\alpha)$  and   $v_{\mu}(m;   k,\alpha)$  denote   the
Rarita-Schwinger  spinors  for $J=3/2$  particles  with momentum  $k$,
helicity $\alpha$ and mass $m$ \cite{ioffe,benmerrouche,hemmert}.
%
Eqs.~(\ref{nkstar}),(\ref{n8k8star})    and    (\ref{diquark-type.SV})
couples to  $J^P=3/2^+$ (anti-)baryons as  well.  In this  case, their
couplings involve $\gamma_5$ in the following way:
\begin{eqnarray}
  \langle 0|
  \psi_{\mu}(0)
  |B_{3/2^+}(k,\alpha)\rangle
  &=&
  \lambda_{3/2^+} \gamma_5 u_{\mu}(m_{3/2^+}; k,\alpha)
  \label{coupling.2}
  \\\nonumber
  \langle 0|
  \bar \psi_{\mu}(0)
  |\bar{B}_{3/2^+}(k,\alpha)\rangle
  &=&
  -
  \lambda_{3/2^+}^*
  \bar v_{\mu}(m_{3/2^+}; k,\alpha) \gamma_5,
\end{eqnarray}
where
$|B_{3/2^+}(k,\alpha)\rangle$  and  $|\bar B_{3/2^+}(k,\alpha)\rangle$
denote $J^P=3/2^+$ (anti-)baryon states with momentum $k$, helicity
$\alpha$ and mass $m_{3/2^+}$.
``$-$''  originates  from  the  anti-commutativity of  $\gamma_0$  and
$\gamma_5$.

To  derive the  spectral  representation,  the best  way  would be  to
express it in the operator representation in the following way:
\begin{equation}
  G_{\mu\nu}(\tau,\vec x)
  =
  Z^{-1}
  \mbox{Tr}
  \left(
  e^{-\beta H}
  T_{\tau}[\psi_{\mu'}(\tau,\vec x) \bar\psi_{\mu}(0)]
  \right),
  \label{correlator.a.1}
\end{equation}
where  $\beta$ denotes the  temporal extent  of the  lattice, $H\equiv
H_{\rm QCD}$ denotes the QCD Hamiltonian,
$Z\equiv \mbox{Tr}(e^{-\beta H})$ denotes the partition function,
and  $T_{\tau}[*]$  represents  the  time-ordered  product  along  the
imaginary time direction.
The interpolating fields are  represented in the Heisenberg picture in
imaginary-time,   i.e.,   $\psi_{\mu}(\tau,\vec   x)  =   e^{\tau   H}
\psi_{\mu}(0,\vec x)  e^{-\tau H}$ and  $\bar\psi_{\mu}(\tau,\vec x) =
e^{\tau H} \bar\psi_{\mu}(0,\vec x) e^{-\tau H}$.
By  restricting  ourselves to  the  interval  $0  \le \tau  <  \beta$,
\Eq{correlator.a.1} reduces to
\begin{equation}
  G_{\mu\nu}(\tau,\vec x)
  =
  \mbox{Tr}
  \left(
  \frac{e^{-\beta H}}{Z}
  \psi_{\mu'}(\tau,\vec x)
  \bar\psi_{\mu}(0)
  \right).
  \label{expression.1}
\end{equation}
Note that it can be equivalently expressed as
\begin{equation}
  G_{\mu\nu}(\tau,\vec x)
  =
  \mbox{Tr}
  \left(
  \psi_{\mu'}(\tau-\beta,\vec x)
  \frac{e^{-\beta H}}{Z}
  \bar\psi_{\mu}(0)
  \right).
  \label{expression.2}
\end{equation}
In the large  $\beta$ limit, we have $e^{-\beta  H}/Z \simeq |0\rangle
\langle   0|$,   which   is   inserted  into   \Eq{expression.1}   and
\Eq{expression.2}.
Note  that   the  resulting  two  expressions   serve  as  independent
contributions  to  the original  ``Tr'',  i.e., \Eq{expression.1}  (or
\Eq{expression.2}).
Hence, we keep these two contributions to obtain
\begin{eqnarray}
  \lefteqn{
    G_{\mu\nu}(\tau,\vec x)
  }
  \\\nonumber
  &\simeq&
  \langle 0 |
  \psi_{\mu'}(\tau,\vec x)
  \bar\psi_{\mu}(0)
  | 0 \rangle
  +
  \langle 0 |
  \bar\psi_{\mu}(0) \psi_{\mu'}(\tau-\beta,\vec x)
  | 0\rangle.
\end{eqnarray}
Note that the 1st term corresponds to the forward propagation, whereas
the 2nd term to the backward propagation.
By  inserting single-(anti-)baryon intermediate  states, and  by using
\Eq{coupling.1} and \Eq{coupling.2}, we are left with
\begin{widetext}
\begin{eqnarray}
  G_{\mu\nu}(\tau,\vec x)
  &=&
  \sum_{\alpha=1}^4
  \int {d^3k\over(2\pi)^3}
  {m_{3/2^+}\over k_0}
  e^{-\tau k_0}
  \langle 0 | \psi_{\mu'}(\vec x) | B_{3/2^+}(k,\alpha)\rangle
  \langle B_{3/2^+}(k,\alpha) | \bar\psi_{\mu}(0)| 0 \rangle
  +
  \cdots.
  \\\nonumber
  &=&
  |\lambda_{3/2^-}|^2
  \int {d^3 k\over (2\pi)^3}
  \frac{m_{3/2^-}}{k_0}
  (-1) P^{(3/2)}_{\mu\nu}(k)
  \left[
  e^{-\tau k_0}
  e^{i\vec k\cdot\vec x}
  \left( \frac{m_{3/2^-} + \fslash{k}}{2 m_{3/2^-}} \right)
  +
  e^{-(\beta - \tau) k_0}
  e^{-i\vec k\cdot\vec x}
  \left( \frac{m_{3/2^-} - \fslash{k}}{2 m_{3/2^-}} \right)
  \right]
  \\\nonumber
  &-&
  |\lambda_{3/2^+}|^2
  \int {d^3 k\over (2\pi)^3}
  \frac{m_{3/2^+}}{k_0}
  (-1)P^{(3/2)}_{\mu\nu}(k)
  \left[
  e^{-\tau k_0}
  e^{i\vec k\cdot\vec x}
  \left( \frac{m_{3/2^+} - \fslash{k}}{2 m_{3/2^+}} \right)
  +
  e^{-(\beta - \tau) k_0}
  e^{-i\vec k\cdot\vec x}
  \left( \frac{m_{3/2^+} + \fslash{k}}{2 m_{3/2^+}} \right)
  \right],
\end{eqnarray}
\end{widetext}
where  $k_0\equiv\sqrt{m_{3/2^-}^2 + \vec  k^2}$ for  $J^P=3/2^-$,
$k_0\equiv\sqrt{m_{3/2^+}^2 + \vec k^2}$ for $J^P=3/2^+$,
and the following identities are used.
\begin{eqnarray}
  \sum_{\alpha=1}^4
  u_{\mu}(m; k,\alpha) \bar u_{\nu}(m; k,\alpha)
  &=&
  - \frac{m + \fslash{k}}{2m}
  P^{(3/2)}_{\mu\nu}(k)
  \\\nonumber
  \sum_{\alpha=1}^4
  v_{\mu}(m; k,\alpha) \bar v_{\nu}(m; k,\alpha)
  &=&
  \frac{m - \fslash{k}}{2m}
  P^{(3/2)}_{\mu\nu}(k),
\end{eqnarray}
where  $P^{(3/2)}_{\mu\nu}(k)$  is the  spin  3/2 projection  operator
defined as
\begin{equation}
  P^{(3/2)}_{\mu\nu}(k)
  \equiv
  g_{\mu\nu} 
  - \frac1{3}\gamma_{\mu}\gamma_{\nu}
  - \frac1{3k^2}
  \left(
  \fslash{k}\gamma_{\mu}k_{\nu}
  + k_{\mu}\gamma_{\nu}\fslash{k}
  \right).
\end{equation}
By  performing  the  integration   over  $\vec  x$  for  zero-momentum
projection, and  by replacing the Minkowskian gamma  matrices by their
Euclidean   counterparts,   we   finally   arrive  at   the   spectral
representation (\Eq{correlator.1} with \Eq{correlator.2}).

The  derivation of  the  spin 1/2  parts  is obtained  by repeating  a
similar  procedure using  the following  parameterizations  instead of
\Eq{coupling.1} and \Eq{coupling.2} as
\begin{eqnarray}
  \lefteqn{
    \langle 0 |
    \psi_{\mu}(0)
    |B_{1/2^-}(k,\alpha)\rangle
  }
  \\\nonumber
  &=&
  \left(
  \lambda_{1/2^-} \gamma_{\mu}
  +
  \lambda'_{1/2^-} k_{\mu}
  \right)
  u(m_{1/2^-}; k,\alpha)
  \\\nonumber
  \lefteqn{
    \langle 0 |
    \bar \psi_{\mu}(0)
    |\bar B_{1/2^-}(k,\alpha)\rangle
  }
  \\\nonumber
  &=&
  \bar v(m_{1/2^-}; k,\alpha)
  \left(
  \lambda_{1/2^-}^* \gamma_{\mu}
  -
  {\lambda'}_{1/2^-}^* k_{\mu}
  \right)
  \\\nonumber
  \lefteqn{
    \langle 0 |
    \psi_{\mu}(0)
    |B_{1/2^+}(k,\alpha)\rangle
  }
  \\\nonumber
  &=&
  \left(
  \lambda_{1/2^+} \gamma_{\mu}
  +
  \lambda'_{1/2^-} k_{\mu}
  \right)
  \gamma_5
  u(m_{1/2^+}; k,\alpha)
  \\\nonumber
  \lefteqn{
    \langle 0 |
    \bar \psi_{\mu}(0)
    |\bar B_{1/2^-}(k,\alpha)\rangle
  }
  \\\nonumber
  &=&
  -
  \bar v(m_{1/2^-}; k,\alpha)
  \gamma_5
  \left(
  \lambda_{1/2^-}^* \gamma_{\mu}
  -
  {\lambda'}_{1/2^-}^* k_{\mu}
  \right),
\end{eqnarray}
where         $|B_{1/2^\pm}(k,\alpha)\rangle$        and        $|\bar
B_{1/2^\pm}(k,\alpha)\rangle$  denote the  $J^P=1/2^\pm$ (anti-)baryon
states with momentum $k$, helicity $\alpha$ and mass $m_{1/2^\pm}$.
$u(m;k,\alpha)$  and $v(m;k,\alpha)$  denote the  Dirac  bispinors for
spin 1/2 particles with mass  $m$, momentum $k$ and helicity $\alpha$.
$\lambda_{1/2^\pm}$  and $\lambda'_{1/2^\pm}$ represent  the couplings
to $J^P=1/2^{\pm}$ (anti-)baryons.

\end{document}